\documentclass[preprint]{elsarticle}
\usepackage{amsmath,amssymb}
\usepackage{bm}
\usepackage{booktabs}
\usepackage{float}
\usepackage{epstopdf}
\usepackage{geometry}
\usepackage{multirow}

\newcommand{\RN}[1]{%
  \textup{\uppercase\expandafter{\romannumeral#1}}%
}

\newcommand\Tstrut{\rule{0pt}{2.6ex}}         

\makeatletter
\def\nfrac#1#2{\check@mathfonts%
    \raise.5ex\hbox{\the\scriptfont0 #1}%
    \kern-.1em/\kern-.15em%
    \lower.25ex\hbox{\the\scriptfont0 #2}}
\makeatother

\biboptions{sort&compress}
\usepackage{hyperref}
\hypersetup{colorlinks=true,breaklinks,linktocpage}

\begin{document}
\begin{frontmatter}
\title{Ductile and brittle crack-tip response in equimolar refractory high-entropy alloys}

\author[KTH,NCSU,WRCP]{Xiaoqing Li\corref{cor}}
\ead{xiaoqli@kth.se}
\address[KTH]{Department of Materials Science and Engineering, KTH - Royal Institute of Technology, Stockholm SE-10044, Sweden}
\address[NCSU]{Department of Materials Science and Engineering, North Carolina State University, Raleigh, North Carolina 27695, USA}
\address[WRCP]{Research Institute for Solid State Physics and Optics, Wigner Research Center for Physics, Budapest H-1525, P.O. Box 49, Hungary}
\author[KTH]{Wei Li}
\author[NCSU]{Douglas L. Irving}
\author[WRCP]{Lajos K. Varga}
\author[KTH,WRCP,UU]{Levente Vitos}
\address[UU]{Department of Physics and Astronomy, Division of Materials Theory, Uppsala University, Box 516, SE-75120, Uppsala, Sweden}
\author[KTH]{Stephan Sch\"onecker\corref{cor}}
\ead{stesch@kth.se}
\cortext[cor]{Corresponding author}

\begin{abstract}
Understanding the strengthening and deformation mechanisms in refractory high-entropy alloys (HEAs), 
proposed as new high-temperature material,
is required for improving their typically insufficient room-temperature ductility.
Here,
density-functional theory simulations and a continuum mechanics analysis were 
conducted to systematically investigate the competition between cleavage 
decohesion and dislocation emission from a crack tip in the body-centered 
cubic refractory HEAs HfNbTiZr, MoNbTaVW,  
MoNbTaW, MoNbTiV, and NbTiVZr.
This crack-tip competition 
is evaluated for tensile loading and a totality of 15 crack 
configurations and slip systems.
Our results predict that dislocation plasticity at the crack tip is generally 
unfavorable -- although the competition is close for some crack orientations,
suggesting intrinsic brittleness and low crack-tip fracture toughness in these five HEAs at zero temperature. 
Fluctuations in local alloy composition, investigated for HfNbTiZr, can locally reduce the resistance to dislocation emission for a slip system relative to the configuration average of that slip system, but do not change the dominant crack-tip response.
In the case of single-crystal MoNbTaW, where an experimental, room-temperature 
fracture-toughness value is available for a crack on a \{100\} plane, 
theoretical and experimental results agree favorably. Factors that may limit the 
agreement are discussed.
We survey the effect of material anisotropy on preferred crack tip 
orientations, which are found to be alloy specific. 
Mixed-mode loadings are found to shift the competition in favor of cleavage or dislocation nucleation, 
depending on crack configuration and amplified by the effect of material anisotropy on crack tip stresses. 
\end{abstract}


\end{frontmatter}

\section{\label{sec:introduction}Introduction}

Increasing efficiency in high-temperature energy production and energy conversion applications 
demands new metallic alloys with superior mechanical properties and high oxidation resistance that can operate at temperatures higher than current Ni-based superalloys. 
As mechanical properties generally degrade at above approximately 0.6 homologous temperature, 
a rationale for exploring metals based on refractory elements is their high-melting points, which may increase the maximum operating temperature.
Consistent with this idea, 
refractory high-entropy alloys (HEAs) have been proposed as novel candidate high-temperature materials~\cite{Senkov:2010,Senkov:2011a,Senkov:2011b,Senkov:2012,Fazakas:2014,Maiti:2016,Wu:2014}. 
Refractory HEAs crystallize in body-centered cubic (bcc) crystal structure and may contain intermetallic phases~\cite{Senkov:2018}.
Some of the refractory HEAs developed so far indeed exhibit impressive strength up to 1600$^\circ$C~\cite{Senkov:2011,Gorsse:2017,Senkov:2018}.
Their typically poor bulk ductility and low fracture toughness at room temperature, however, make processing and machining difficult, and limit current application. 
These challenges motivate further research into tough, high-specific strength, and ductile refractory HEAs.

A major challenge in understanding metallic alloys' mechanical properties is that metallic materials are inherently multiscale, with macroscopic mechanical response (strength, ductility, toughness, etc.) determined by details of the microstructure and dependent on intrinsic and extrinsic materials properties as well as external conditions~\cite{Pineau:2016}.
Nonetheless, a widely accepted requirement for achieving high fracture toughness is that the material is intrinsically ductile. Intrinsic ductility in crystalline materials is closely related to the capability for dislocation plasticity near crack tips.
An intrinsically ductile material may be distinguished from an intrinsically brittle one in terms of the atomic structure at the tip of a sharp crack, corresponding to two major competing modes of response~\cite{Thomson:1986}: 
the material is said to be intrinsically ductile (or brittle) if external loading will blunt the crack by dislocation emission rather than 
cleaving by crack propagation (or vice versa).

Theoretical analysis of this competition is well established~\cite{Kelly:1967,Rice:1974,Thomson:1986,Rice:1992,Sun:1994,Rice:1994,Zhou:1994,Xu:1995,Fischer:2001,Andric:2017}. 
Kelly \emph{et al.}~\cite{Kelly:1967} and later Rice and Thomson~\cite{Rice:1974} developed continuum models to predict how a pre-cracked body responds to applied load. The critical loads for dislocation emission and crack propagation, characterized by stress intensity factors or energy release rates, were compared and the event requiring the lower threshold was predicted to dominate. 
Rice~\cite{Rice:1992} greatly advanced the viewpoint of brittle versus ductile fracture behavior by incorporating the Peierls framework into the dislocation nucleation description. This analysis introduced the unstable stacking fault (USF) energy $\gamma_{\text{usf}}$ as the key material parameter measuring the resistance to dislocation nucleation. 
The Rice-Thomson model and the Rice framework have been broadened over the years~\cite{Sun:1994,Rice:1994,Zhou:1994,Xu:1995,Fischer:2001,Andric:2017}, for instance, to account for elastic anisotropy, the effect of crack blunting, three-dimensional dislocation nuclei, successive nucleation events, and surface ledges formed at the crack tip.
Evidently, a comprehensive understanding of these processes must involve some degree of atomistic modeling (for example, Refs.~\cite{Zhu:2004,Gouriet:2012,Wu:2015} yielded important results pertaining to embryonic dislocation loop emission and crack blunting), last but not least to reconcile continuum models with results of atomistic modeling.
Nevertheless, atomistic modeling necessitates an underlying interatomic potential that reproduces the necessary material parameters, dislocation cores,  fracture properties, etc., accurately. 

In this work, we report a systematic theoretical assessment of the intrinsic ductility of some bcc refractory HEAs by studying the competition between dislocation emission and cleavage at a crack tip, which is not available yet. 
Further impetus to this investigation is due to recent fracture experiments for pre-cracked, small-sized, single-crystal MoNbTaW specimens~\cite{Zou:2017}, which suggested a quasi-cleavage fracture mode with limited amount of crack tip plasticity.
We investigate the crack-tip competition in five, previously synthesized, equimolar HEAs for tensile loading and 15 crack configurations and slip systems.
Our theoretical predictions build on the analytical solutions of the Rice theory~\cite{Rice:1992,Sun:1994} for the nucleation of two-dimensional dislocation geometries. The analysis draws upon continuum elasticity to determine the stress field for a crack in an infinite, anisotropic linear elastic medium, similar to previous work for Mg~\cite{Wu:2015}. 
The necessary material parameters are determined by quantum-mechanical calculations based on density-functional theory (DFT) and first-principles alloy theory. 
Targeting the intrinsic properties of the bcc matrix is essential in understanding the strengthening and deformation mechanisms in refractory HEAs and to provide guidance on improving the ductility of the bcc matrix. 
The current approach complements previously proposed ductilization strategies for refractory HEAs, which focused at, for instance, grain boundaries and size effects~\cite{Zou:2014,Zou:2017}, compositional effects on the tensile ductility of polycrystalline bulk~\cite{Sheikh:2016}, and strain-induced or stress-induced phase transformations~\cite{Lilensten:2017,Huang:2017b}.

The remainder of this paper is organized as follows. In Sec.~\ref{sec:computationalmethod},
we give a brief exposition of the Rice theory and the employed criteria for brittle crack growth and dislocation emission, and detail all relevant methodological and computational aspects. Section~\ref{sec:results} presents the main results and discussion. We compare our findings to available experimental results and discuss factors that may limit the agreement. 
We show how deviation from pure tensile loading affects the competition and study the effect of material anisotropy. 
We investigate the effect of fluctuations in local alloy composition on the crack-tip competition for HfNbTiZr.
We comment on surface ledges, crack-tip twinning, and the effect of finite-temperatures.
Section~\ref{sec:Conclusions} concludes.

\section{\label{sec:computationalmethod}Methodological details and computational method}

\subsection{Criteria for crack extension and dislocation nucleation at a crack tip}

\begin{figure}
 \begin{center}
 \resizebox{0.65\linewidth}{!}{\includegraphics[clip]{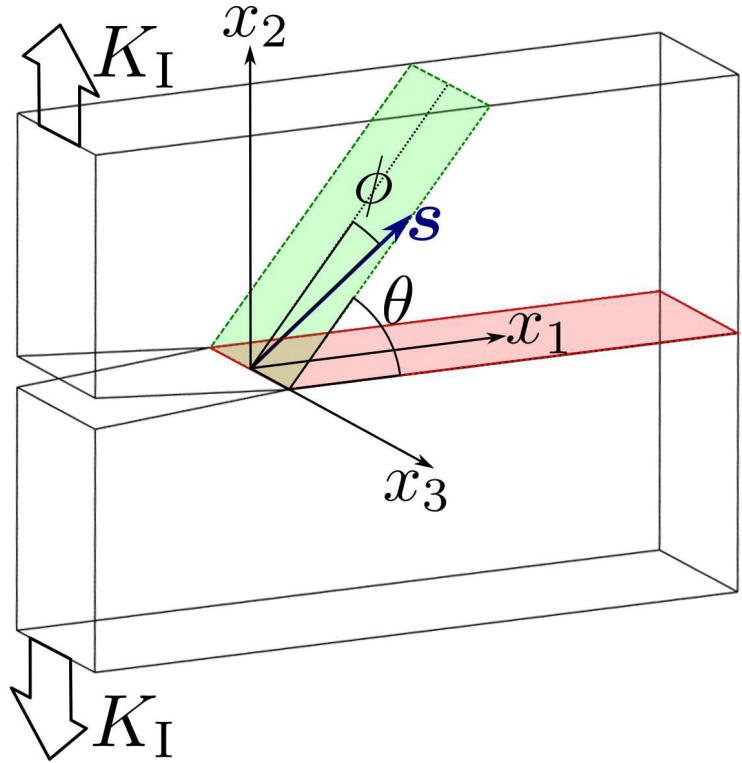}}
 \caption{\label{fig:geometry}Schematic of a two-dimensional planar crack in an elastic media subject to tensile $K_\RN{1}$ loading.
 The intersection line between the crack plane (in light red) and the slip plane (in light green) is the crack front. The slip direction is along ${\bm s}$. The crack indicated by a wedge is atomically sharp and treated as slit crack in the LEFM analysis (\ref{sec:app}), where the origin of the crack coordinate system $(x_1,x_2,x_3)$ is shifted so that the crack in centrally located at $x_1=0$.}
 \end{center}
\end{figure}

Specifically focusing on tensile (or mode $\RN{1}$) loading, we consider two competing mechanisms, crack propagation due to Griffith cleavage~\cite{Griffith:1921} and dislocation emission from a sharp crack tip within the Peierls framework~\cite{Rice:1992,Sun:1994}. 
We resort to linear elastic fracture mechanics (LEFM) for anisotropic media to determine the displacement and stress fields (in plane strain) for a sharp crack in an infinite medium, which is summarized in~\ref{sec:app} and referred to where appropriate.
In combination with first-principles derived material parameters, the conditions introduced below allow determining the critical stress intensity factors for crack propagation $K_{\RN{1}\text{c}}$ and dislocation emission from the crack tip $K_{\RN{1}\text{e}}$.

\paragraph{Geometry}
Two different rectilinear coordinate systems are employed: Cartesian coordinates $(x_1,x_2,x_3)$ and cylindrical-polar coordinates ($r$, $\theta$, $x_3$).
A slip plane intersects the crack plane and the intersection line is the crack front; see Fig.~\ref{fig:geometry}.
$x_1$ is the crack tip (or crack extension) direction, $x_2$ is the normal to the crack plane, and $x_3$ is along the crack front direction.
 The slip plane makes an angle $\theta$ with the crack plane and the slip direction ${\bm s}$ makes an angle $\phi$ with the crack tip direction in the slip plane. 
The slip direction is defined parallel to the Burgers vector, which in the slip plane has the unit vector ${\bm s}(\phi) = [\cos\phi, 0, \sin\phi]^T$.

\paragraph{Crack extension}
The Irwin energy release rate $G$ of a loaded, sharp crack tip with respect to crack propagation (in original direction, $\theta =0$) in an anisotropic, linear elastic medium under plane strain deformation reads~\cite{Stroh:1958,Barnett:1972},
\begin{align}
 G &= {\bm K}^T {\bm \Lambda } {\bm K}.
 \label{eq:defG}
\end{align}
Here,
${\bm K} = [K_1, K_2, K_3]^T  = [K_\RN{2}, K_\RN{1}, K_\RN{3}]^T$ is the external loading. 
The $K_{\alpha}$ ($\alpha = 1\ldots 3$) are the stress intensity factors; $K_\RN{1}$, $K_\RN{2}$, and $K_\RN{3}$ apply to tensile loading mode, in-plane shear loading mode, and anti-plane shear loading mode, respectively.
${\bm \Lambda } = {\bm L}^{-1}/2$ is the relevant matrix from the Stroh formalism determined by elastic constants $C_{ijkl}$ for the chosen crack configuration; see~\ref{sec:app}.
The Griffith condition for tensile crack growth of an ideally brittle material in crack extension direction under mode $\RN{1}$ loading reads~\cite{Griffith:1921},
\begin{align}
 G &= 2\gamma_{\text{s}},
 \label{eq:GriffithG}
\end{align}
where $\gamma_{\text{s}}$ is the surface energy of the crack plane.
Equating the previous two expression, one finds the critical stress intensity factor for crack propagation under pure $K_\RN{1}$ loading,
\begin{align}
 K_{\RN{1}\text{c}} &= \sqrt{ \frac{ 2\gamma_{\text{s}}}{\Lambda_{22}}}.  
 \label{eq:help1}
\end{align}

\paragraph{Dislocation nucleation}
Rice~\cite{Rice:1992} analyzed dislocation emission from a crack tip within the Peierls framework in isotropic, linear elastic media. Similar to the Peierls-Nabarro model of a dislocation core, Rice's model assumes that the displacement field across the slip plane emanating from the crack tip follows a periodic relation between shear stress and atomic displacement. As load is applied gradually, a slip displacement occurs, which at the point of maximum lattice potential causes an instability of an incipient dislocation configuration, i.e., dislocation nucleation takes place. 
Using the $J$-integral method and treating the slip process zone as a pure shear process, i.e., no tension-shear coupling, Rice derived that the instability for a complete lattice dislocation occurs when
\begin{align}
 G&=\gamma_{\text{usf}}.
 \label{eq:Rice}
\end{align}
That is,
the energy release rate equals the USF energy $\gamma_{\text{usf}}$.  The latter coincides with the maximum lattice potential and may be obtained from a block-like shear displacement of two parts of the crystal along the displacement vector~\cite{Vitek:1968}.
In deriving the previous result, it is assumed that slip is constrained to a straight slip path ${\bm s}$ coinciding with observed slip directions~\cite{Rice:1992} without involving dilational displacement perpendicular to the slip plane, i.e., $\gamma_{\text{usf}}$ in Eq.~\eqref{eq:Rice} is the unrelaxed USF energy. 
A here employed approximate method that accounts for tension effects is to use the relaxed USF energy instead of the unrelaxed value~\cite{Rice:1992,Sun:1993}.

The analysis of dislocation emission from a crack tip was subsequently generalized to anisotropic, linear elastic media by Sun and Beltz~\cite{Sun:1994}. 
The nucleation condition for a general, straight dislocation in the case of \emph{coplanar} crack and slip planes reads~\cite{Sun:1994},
\begin{align}
 \frac{ ( {\bm s}(\phi) {\bm K}_{\text{e}})^2 }{{\bm s}^T(\phi) {\bm \Lambda}^{-1} {\bm s}(\phi)} &= \gamma_{\text{usf}} \qquad(\theta = 0).
 \label{eq:SBinplane}
\end{align}
For pure mode $\RN{2}$ loading the emerging dislocation is of edge type when the slip direction is in the $x_1$ direction ($\phi = 0$);  for pure mode $\RN{3}$ loading, the emerging dislocation is of screw type when the slip direction is in the $x_3$ direction ($\phi = \pi/2$).
The term ${\bm s}(\phi) {\bm K}_\text{e}$ may thus be interpreted as a superposition of these two stress intensity factors for the resolved shear stress along ${\bm s}$. 
The denominator on the left hand side relates ${\bm \Lambda}^{-1}$ to the slip direction.

For a tilted slip plane, the ``exact'' solution for the nucleation condition requires numerical methods. Rice devised an approximate, though justified, projection method~\cite{Rice:1992}, according to which the nucleation criterion for anisotropic, linear elastic media  and under a general set of intensity factors ${\bm K}$ reads~\cite{Sun:1994},
\begin{align}
 \frac{ ( {\bm s}(\phi) {\bm K}^{\text{eff}}_{\text{e}} )^2} {{\bm s}^T(\phi) {\bm \Lambda}^{(\theta)-1} {\bm s}(\phi)}  &= \gamma_{\text{usf}} \qquad (\theta \ne 0).
 \label{eq:SBfinal}
\end{align}
This condition adopts effective stress intensity factors $K^{\text{eff}}_\alpha$, defined through the singular in-plane and anti-plane stress components $\sigma_{\theta \alpha}= K^{\text{eff}}_\alpha /\sqrt{2\pi r}$ acting in the slip plane. They are related to the stress intensity factors acting on the main crack by 
$K^{\text{eff}}_\alpha= F_{\alpha\beta}(\theta) K_\beta$, where the $F_{\alpha\beta}$ contain the $\theta$-dependence of $\sigma_{\theta \alpha}$ for $K_\beta$. The $F_{\alpha\beta}$ are obtained from the elastic stress field for the crack geometry; see~\ref{sec:app}.
The denominator on the left hand side  and
the matrix ${\bm \Lambda}^{(\theta)-1} = {\bm Q}(\theta) {\bm \Lambda}^{-1} {\bm Q}^T(\theta)$ relate ${\bm \Lambda}^{-1}$  from the main crack to the slip plane and slip direction.
${\bm Q}$ is the three-dimensional rotation matrix around $x_3$ through angle $\theta$.

Specifically, pure mode $\RN{1}$ loading on the main crack induces shear stresses in the slip plane, for which the critical
stress intensity factor for dislocation emission simplifies to
\begin{align}
 K_{\RN{1}\text{e}} &= \frac{\sqrt{\gamma_{\text{usf}}\, {\bm s}^T(\phi) {\bm \Lambda}^{(\theta)-1} {\bm s}(\phi)}}{F_{12}(\theta) \cos\phi  + F_{32}(\theta) \sin\phi }.
 \label{eq:SBfinalKI}
\end{align}

The material parameters required for this analysis are the tensor of elastic moduli, the surface energy of the crack plane, and the USF energy of the slip plane in the slip direction.

\subsection{Crack configurations and slip systems}

Several crack configurations were examined since the USF energies of different slip systems in bcc metals are well known to be different~\cite{Vitek:1968}, indicating that the crack-tip competition is orientation dependent.
We considered two commonly active slip systems in bcc metals, $\nfrac{1}{2}\langle 111 \rangle \{110\}$ and $\nfrac{1}{2}\langle 111 \rangle \{112\}$, where $\nfrac{1}{2}\langle 111 \rangle$ is the Burgers vector.
We selected the $\{100\}$ and $\{110\}$ crystallographic planes as crack planes, as they are the typically favored and observed cleavage planes in bcc materials~\cite{Tyson:1973}, in addition to considering $\{111\}$ as an example for a more corrugated crack plane.

\begin{figure}[th]
	\begin{center}
		\begin{tabular}{@{}ll@{}}
		(a) model 1 & (b) model 4 \\
			\resizebox{0.45\linewidth}{!}{\includegraphics[clip]{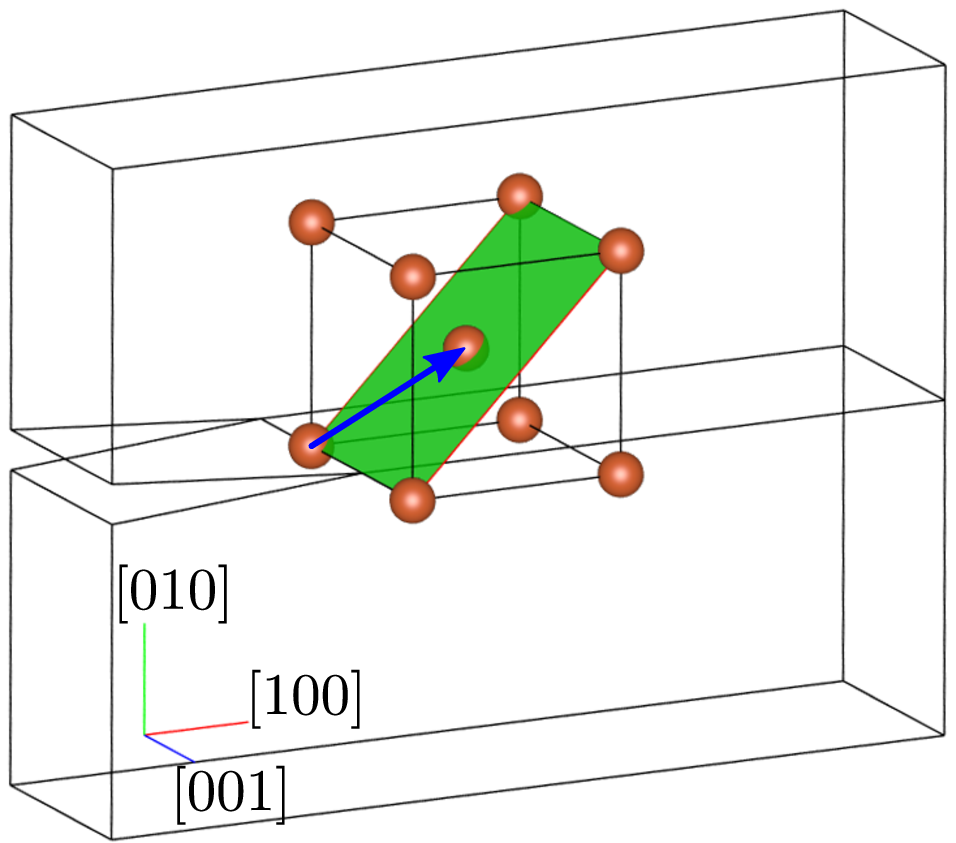}}&
			\resizebox{0.45\linewidth}{!}{\includegraphics[clip]{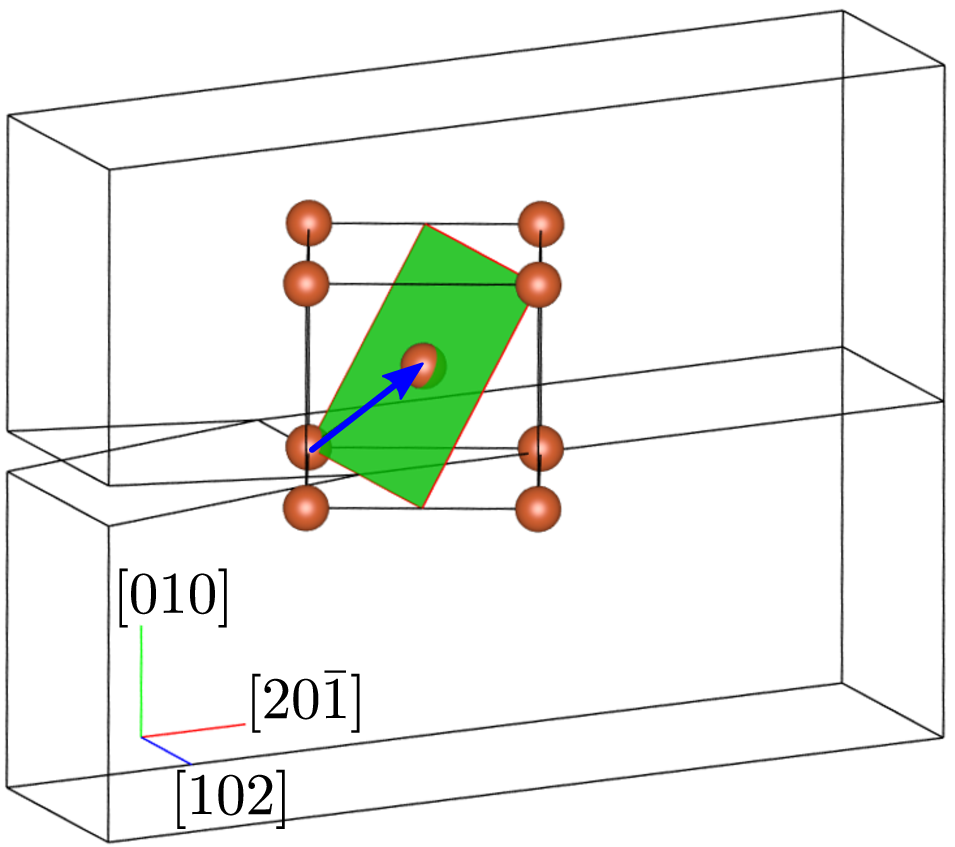}}\\[5pt]
		(c) model 8 & (d) model 13 \\	
			\resizebox{0.45\linewidth}{!}{\includegraphics[clip]{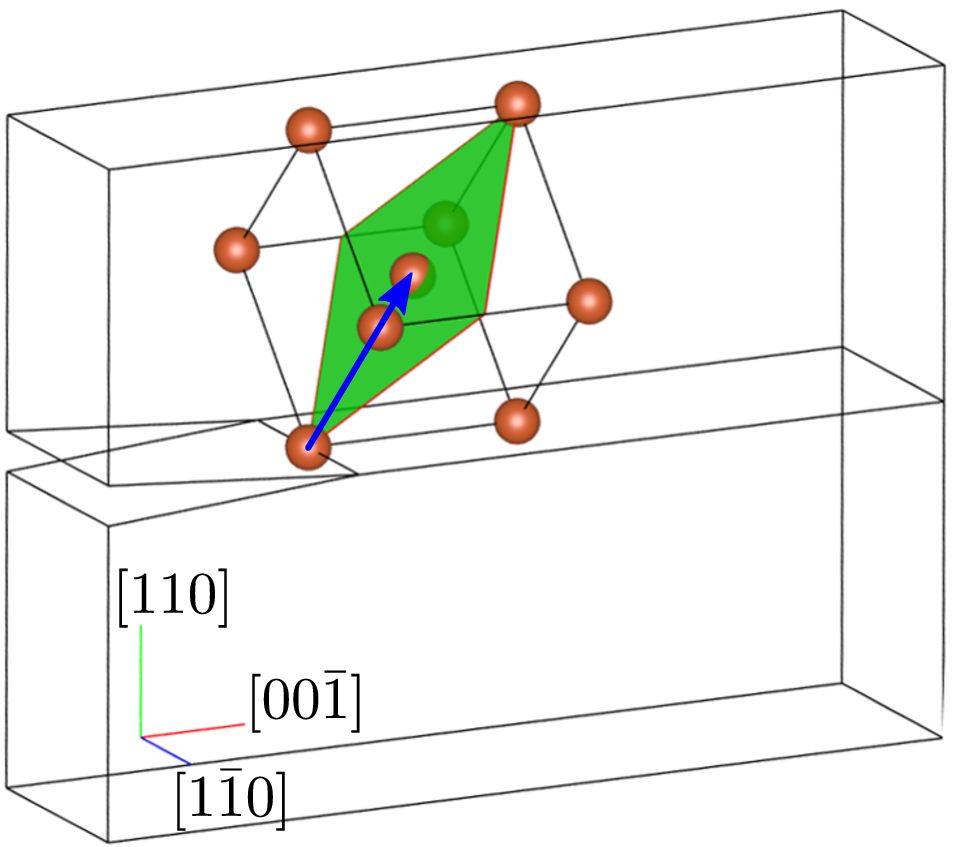}}&
			\resizebox{0.45\linewidth}{!}{\includegraphics[clip]{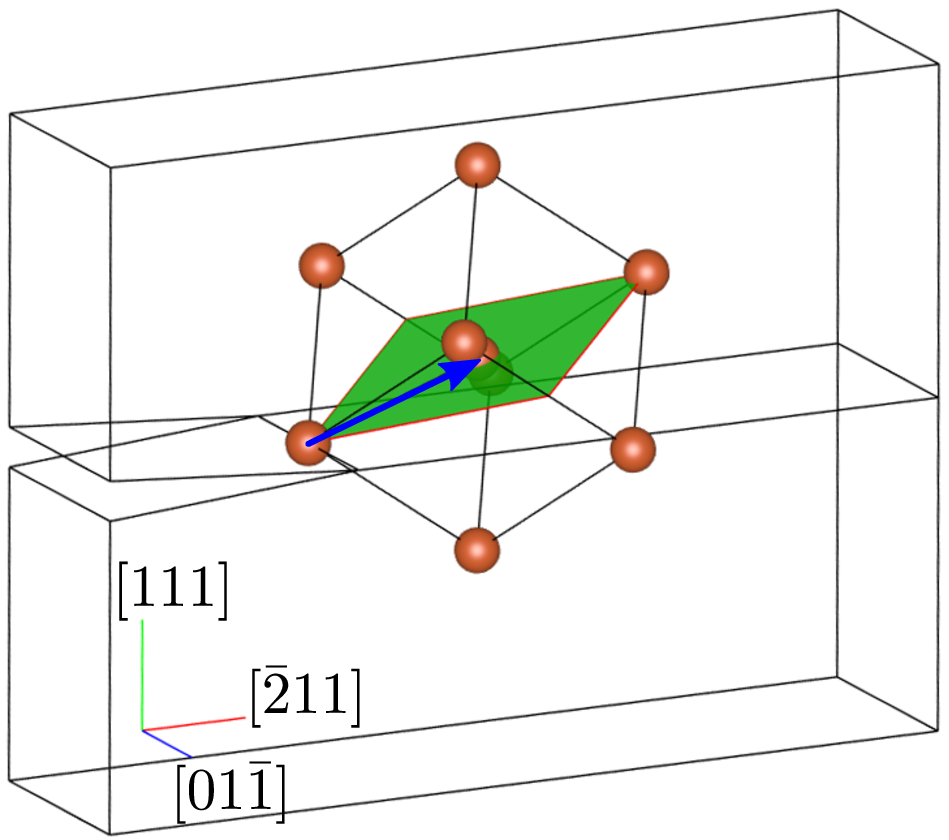}}\\	
		\end{tabular}
		\caption{\label{fig:diagram}Schematic diagrams of four selected crack configurations and dislocation slip systems: (a) $\{010\}$ crack plane and $\{110\}$ slip plane, (b)  $\{010\}$ crack plane and $\{112\}$ slip plane, (c) $\{110\}$ crack plane, and $\{112\}$ slip plane, (d) $\{111\}$ crack plane, and $\{112\}$ slip plane. Depicted is a single conventional bcc unit cell, the slip plane intersecting the unit cell in green, and the $\nfrac{1}{2}\langle 111 \rangle$ Burgers vector in blue.}
	\end{center}
\end{figure}

The totality of considered, distinct crack configurations and slip systems available for dislocation emission in mode $\RN{1}$ loading are listed in Table~\ref{table:orientation} (among equivalent one, those with $\theta \in (0,90^\circ]$ and $\phi \in [0,90^\circ]$ were selected).
Coplanar crack and slip planes are not considered in the present application, as the in-plane shear $\sigma_{\theta r}(\theta = 0)$ is identical zero under pure mode $\RN{1}$ loading. 
Each crack model is labeled for convenience: models 1-4 have the (010) crack plane, models 5-9 have the (110) crack plane, and the remaining models the (111) crack plane. 
Schematic diagrams of four chosen crack configurations (models 1, 4, 8, and 13) are exemplified in Fig.~\ref{fig:diagram}.

\begin{table}[tbh]
        \caption{\label{table:orientation}Summary of distinct crack configurations and slip systems in bcc crystals considered in mode $\RN{1}$ loadings. The crystallographic orientation of the crack is denoted by $x_1$-$x_2$-$x_3$.}
        \begin{tabular}{cllrr}
                \toprule
                Model & Crack orientation & Slip system & $\theta$ ($^\circ$) & $\phi$ ($^\circ$)\\
                \midrule
                1 &[100]-[010]-[001]& $\nfrac{1}{2}[111](\bar{1}10)$& 45.00 & 35.26\\
                2 &[10$\bar{1}]$-[010]-[101] & $\nfrac{1}{2}[111](\bar{1}01)$& 90.00& 54.74 \\
                3 &[101]-[010]-[$\bar{1}01$] & $\nfrac{1}{2}[111](\bar{1}2\bar{1})$& 35.00& 0.00 \\
                4 &[20$\bar{1}]$-[010]-[102] & $\nfrac{1}{2}[111](\bar{2}11)$& 65.00& 50.77\\
                \hline
                5 &[1$\bar{1}\bar{2}]$-[110]-[1$\bar{1}1]$ &$\nfrac{1}{2}[1\bar{1}1]$(011)& 60.00 & 0.00\Tstrut\\
                6 &[$\bar{1}1\bar{2}]$-[110]-[$1\bar{1}\bar{1}]$ & $\nfrac{1}{2}[11\bar{1}](101)$&60.00 & 19.47 \\
                7 &$[1\bar{1}0]$-$[110]$-$[001]$ & $\nfrac{1}{2}[111](\bar{1}10)$&90.00 & 35.26\\
                8 &[00$\bar{1}]$-[110]-[1$\bar{1}0]$ &$\nfrac{1}{2}[11\bar{1}](112)$&54.74& 0.00\\
                9 &[3$\bar{3}2]$-[110]-[$\bar{1}13]$ & $\nfrac{1}{2}[111](\bar{1}2\bar{1})$&73.22& 31.48 \\
                \hline
                10 &[$\bar{1}\bar{1}2]$-[111]-[$\bar{1}10]$ & $\nfrac{1}{2}[\bar{1}11]$(110)&35.26& 54.74\Tstrut\\
                11 &[1$\bar{1}0]$-[111]-[$\bar{1}\bar{1}2]$ & $\nfrac{1}{2}[111](\bar{1}10)$&90.00& 0.00\\
                12 &[1$\bar{1}0]$-[111]-[$\bar{1}\bar{1}2]$ & $\nfrac{1}{2}[11\bar{1}](\bar{1}10)$&90.00& 70.53\\
                13&[$\bar{2}11]$-[111]-[01$\bar{1}]$ & $\nfrac{1}{2}[\bar{1}11]$(211)&19.47& 0.00 \\
                14&[2$\bar{1}\bar{1}]$-[111]-[0$\bar{1}1]$ & $\nfrac{1}{2}[111](\bar{2}11)$&90.00& 0.00 \\
                15&[5$\bar{1}\bar{4}]$-[111]-[1$\bar{3}2]$ & $\nfrac{1}{2}[1\bar{1}1](\bar{1}12)$&61.87& 67.79\\
                \bottomrule
        \end{tabular}
\end{table}

\subsection{Details of the density-functional theory calculations}

DFT simulations were employed to determine the tensor of elastic moduli, the surface energies, and the USF energies for the refractory HEAs in the bcc structure and in theoretical equilibrium.
The three single-crystal elastic constants (in contracted notation), $C_{11}$, $C_{12}$, and $C_{44}$, were determined by means of mapping the strain energy for prescribed, volume-conserving strains as elaborated in Ref.~\cite{Li:2012}.
The surface energies and USF energies were extracted from slab calculations and tilted super cell calculations with periodic boundary conditions~\cite{Rodney:2017}, respectively.
In both cases, the total energy of the defect-free, bulk reference system was obtained by way of increasing the thickness of the slab/super cell~\cite{Fiorentini:1996} (the increase of the slab/super cell total energy with thickness equals integer multiples of the bulk energy for sufficiently large slab/super cell sizes). Here we derived the bulk energy from a linear fit to three different slab/super cell sizes in order to achieve high numerical stability.
We found that slabs containing 11, 13, and 15 atomic layers with free surfaces separated by vacuum corresponding to seven atomic layers yield converged $\{100\}$ surface
energies, whereas 7, 9, and 11 atomic layers resulted in converged $\{110\}$ surface energies. For the $\{111\}$ termination, we used
6, 12, and 18 atomic layers and vacuum corresponding to six layers (the increments differ due to stacking sequence).  
The effect of surface relaxation, considered for the interlayer spacing between the surface layer and the subsurface layer, was evaluated by comparing the energy of the relaxed surface slabs with 13, 12, and 9 atomic layers for the $\{100\}$, $\{110\}$, and $\{111\}$ terminations, respectively, to perfectly truncated bulk slabs with the same thicknesses.
Super cells for the USF configuration were based on 16, 20, and 24 atomic layers for the $\{110\}$ slip planes, and 12, 18, and 24 atomic layers were adopted for the $\{112\}$ slip planes. The relaxation of the interplanar distances perpendicular to the slip plane was considered locally for the two layers forming the fault, and the relaxed USF energies were obtained for the super cell with 16 and 18 layers for $\{110\}$ and $\{112\}$ slip planes, respectively.

The DFT simulations were carried out with the exact muffin-tin orbitals (EMTO) method~\cite{EMTO:1,EMTO:3,cpa:4}. The local-density approximation  parameterized by Perdew and Wang~\cite{Perdew:1992} was adopted to describe exchange and correlation and the total energy was computed via the full charge-density technique~\cite{Vitos:1997b,cpa:4}.
The basis set included $s$, $p$, $d$, and $f$ states. 
The electronic structure of the substitutionally disordered refractory HEAs was computed in the coherent-potential approximation (CPA)~\cite{Soven:1967,cpa:1,cpa:3}, wherein the random alloy crystal potential is replaced by a self-consistently determined effective medium having the translational symmetry of the underlying lattice and yielding configuration-averaged total energies.
The present study does not account for possible short-range order or local lattice relaxation effects, and all results are valid for random solid solutions on a perfect, bcc crystal lattice.
While some of the refractory HEAs considered were predicted~\cite{Ikeda:2019}  to exhibit complex and competing chemical ordering modes (short-range order and long-range order), investigating these effects on the material parameters required is beyond the scope of this study.

Brillouin zone integrations were performed on a $27 \times 27 \times 27$ $k$-points mesh for the elastic constants calculations. For the determination of the planar fault energies, $17 \times 25 \times 1$ $k$-points were found to yield converged
surface energies for the bcc $\{100\}$, $\{110\}$, and $\{111\}$ surface facets, whereas $9 \times 23 \times 2$ and $13 \times 25 \times 2$ $k$-points were used for the $\{110\}$ and $\{112\}$ USFs calculations, respectively. 

An investigation of the effect of fluctuations in local alloy composition on the crack-tip competition was carried out for HfNbTiZr. To this end, we determined the local-environment dependence of USF energies and surface energies using the super cell approach to represent a substitutionally disordered alloy. All further details of these calculations are summarized in~\ref{sec:appLF}.

The accuracy of the EMTO method and the CPA for the elastic properties
and planar fault energies of random alloys including HEAs was demonstrated previously~\cite{Wang:2015,Li:2016,Huang:2018,Xiaoqing:2018}.

\section{\label{sec:results}Results and discussion}

\begin{table*}[tbh]
	\caption{\label{table:lattice_constants}Material parameters of the five bcc HEAs, arranged in order of increasing VEC, used in the LEFM analysis: the elastic constants (in units of GPa), and the relaxed surface energies and USF energies for selected fault planes (in units of J/m$^{2}$). The Zener anisotropy ratio is also listed.}
	\begin{tabular}{@{}lcccccccc@{}}
		\toprule
		HEA & VEC& $C_{11}$& $C_{12}$& $C_{44}$ & $A_\text{Z} $& Plane& $\gamma_{\text{s}}$& $\gamma_{\text{usf}}$ \\[2pt]
		\midrule
		HfNbTiZr &4.25&154.3&127.9&56.7& 4.30 & (010)&1.959& -\\
		&&&&&&(110)&1.960& 0.576\\
		&&&&&&(111)&2.284& -\\
		&&&&&&(112)&-&0.594 \\
		NbTiVZr &4.5&192.8&127.9&51.7& 1.59 &(010)&2.141& -\\
		&&&&&&(110)&2.127&0.701\\
		&&&&&&(111)&2.457&-\\
		&&&&&&(112)& - &0.759\\
		MoNbTiV &5.0&294.7&152.0&48.7& 0.68 &(010)&2.800& -\\
		&&&&&&(110)&2.554&1.077 \\
		&&&&&&(111)&2.935& -\\
		&&&&&&(112)&-&1.069 \\
		MoNbTaVW &5.4&370.8&178.2&51.6& 0.54 &(010)&3.477& -\\
		&&&&&&(110)&2.909&1.448 \\
		&&&&&&(111)&3.372& -\\
		&&&&&&(112)&-&1.447 \\
		MoNbTaW &5.5&411.8&184.6&71.0& 0.63 &(010)&3.667& -\\
		&&&&&&(110)&2.943&1.584 \\
		&&&&&&(111)&3.444& -\\
		&&&&&&(112)&-&1.602 \\
		\bottomrule
	\end{tabular}
\end{table*}

\subsection{\label{sec:parameter}Material parameters}
\begin{figure*}[htb]
	\begin{center}
		\begin{tabular}{@{}ll@{}}
			(a) & (b) \\
			\resizebox{0.45\linewidth}{!}{\includegraphics[clip]{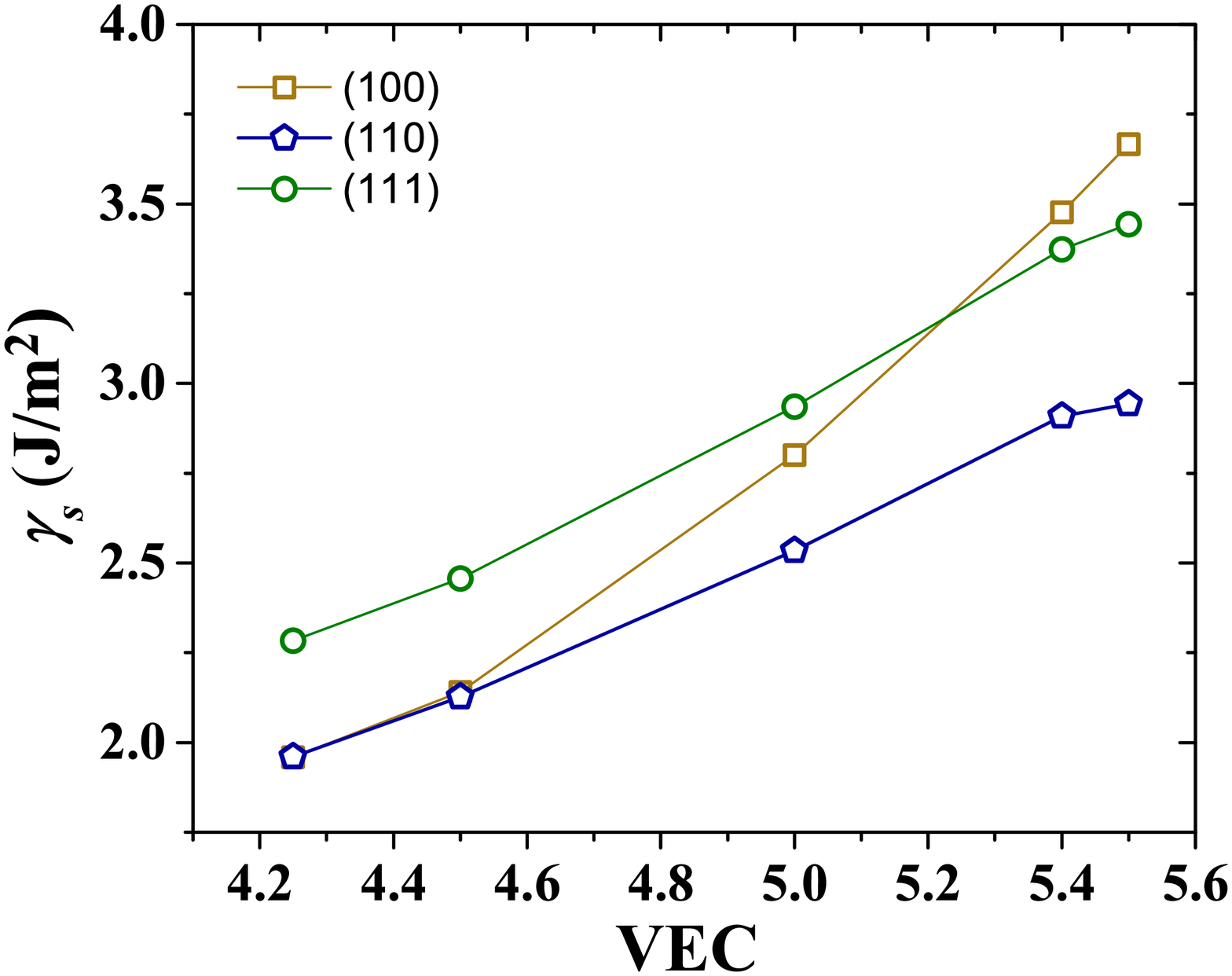}}&
			\resizebox{0.45\linewidth}{!}{\includegraphics[clip]{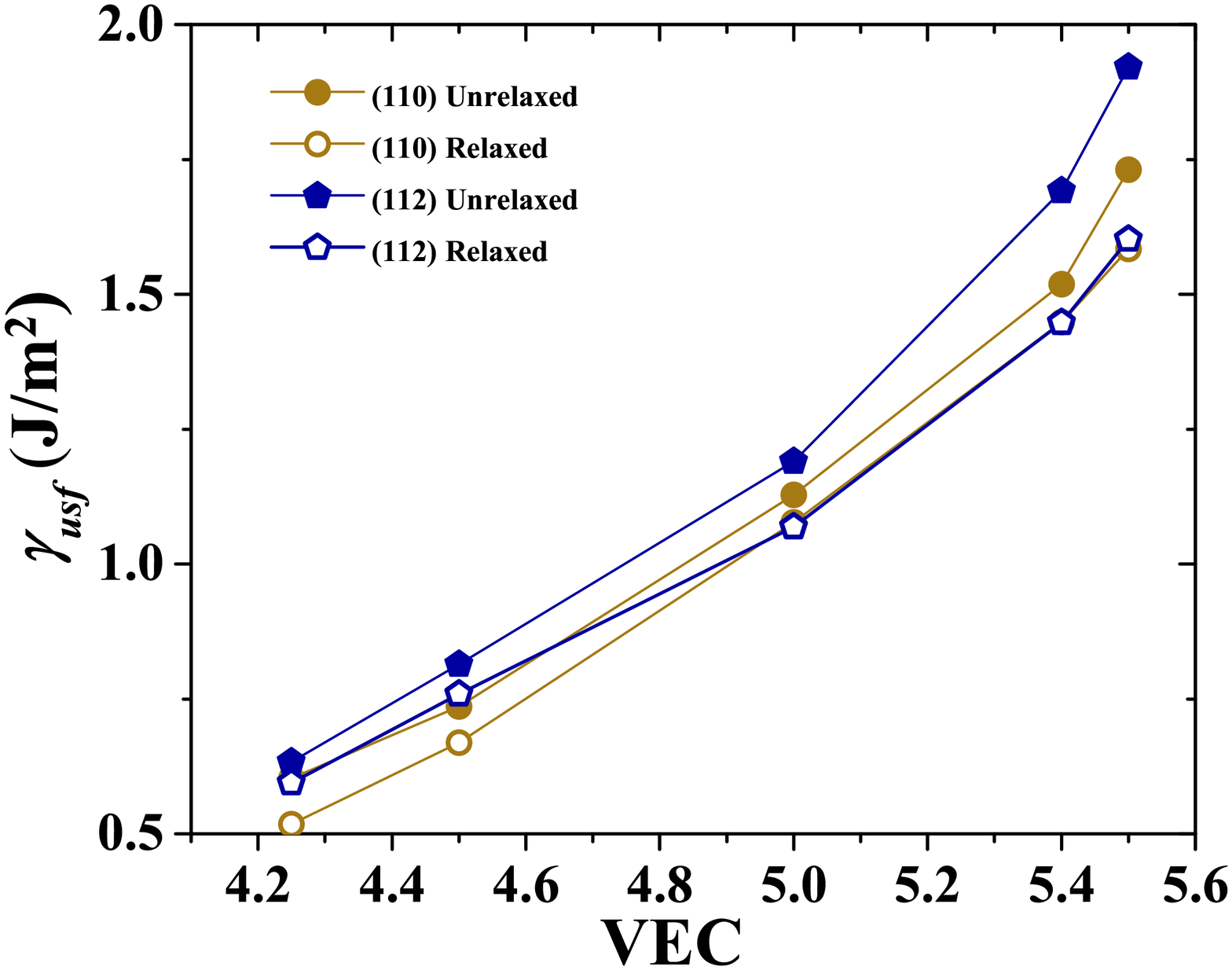}} \\
		\end{tabular}
		\caption{\label{fig:energy}The (a) surface energies (relaxed) and (b) USF energies (relaxed and unrelaxed) of the refractory HEAs as a function of the VEC. Lines guide the eye.}
	\end{center}
\end{figure*}

The computed material parameters used in the LEFM analysis, i.e., the complete set of cubic single-crystal elastic constants and the planar fault energies for the considered crack and slip planes, are listed in Table~\ref{table:lattice_constants} for the  bcc HEAs considered, HfNbTiZr, MoNbTaVW, MoNbTaW, MoNbTiV, and NbTiVZr.
These and the following results are arranged in increasing order of the alloys' average valence electron concentration (VEC, in units of electrons/atom).
The Zener anisotropy ratio $A_{\text{Z}} = 2C_{44} / (C_{11} - C_{12})$ characterizing elastic anisotropy is tabulated alongside in order to support the discussion.

We briefly discuss the surface energies and USF energies due to their simple relation to $K_{\RN{1}} \propto \sqrt{\gamma}$.
The relaxed values for both planar faults are shown in Fig.~\ref{fig:energy} along with the unrelaxed USF energies.
It follows that the surface energies of the three chosen facets increase with the VEC. 
The energy of the $\{100\}$ surface facet shows a stronger dependence on the VEC compared to those of the $\{110\}$ and $\{111\}$ facets. 
We found that the relaxation of the surface geometry lowers the surface energy of the alloys considered in the range of
5-6\%,  5-7\%, and 3-8\% for the $\{100\}$, $\{110\}$, and $\{111\}$ facets, respectively, where the ranges express the influence of relaxation over the five considered alloys (unrelaxed surface energies not shown). 
The USF energies display a similar dependence on the VEC as the surface energy. However, the VEC has a stronger effect on $\gamma_{\text{usf}}$, e.g., $\gamma_{\text{usf}}$ of MoNbTaW with a VEC of 5.0 is three times larger than that of HfNbTiZr with a VEC of 4.25. 
The unrelaxed USF energies are approximately 4-14\,\% and 6-17\,\% larger for the $\langle 111\rangle\{110\}$ and $\langle 111\rangle\{112\}$ slip systems, respectively, than the relaxed values, where again the ranges express the influence of relaxation over the alloys considered.
That is, accounting for tension-shear coupling by means of relaxing the USF energies reduces the critical loading for dislocation emission in the range of 2-8\,\%.

All following results were obtained with the relaxed values of the USF energies.

\begin{figure}[thb]
	\begin{center}
		\resizebox{!}{.5\textheight}{\includegraphics[clip]{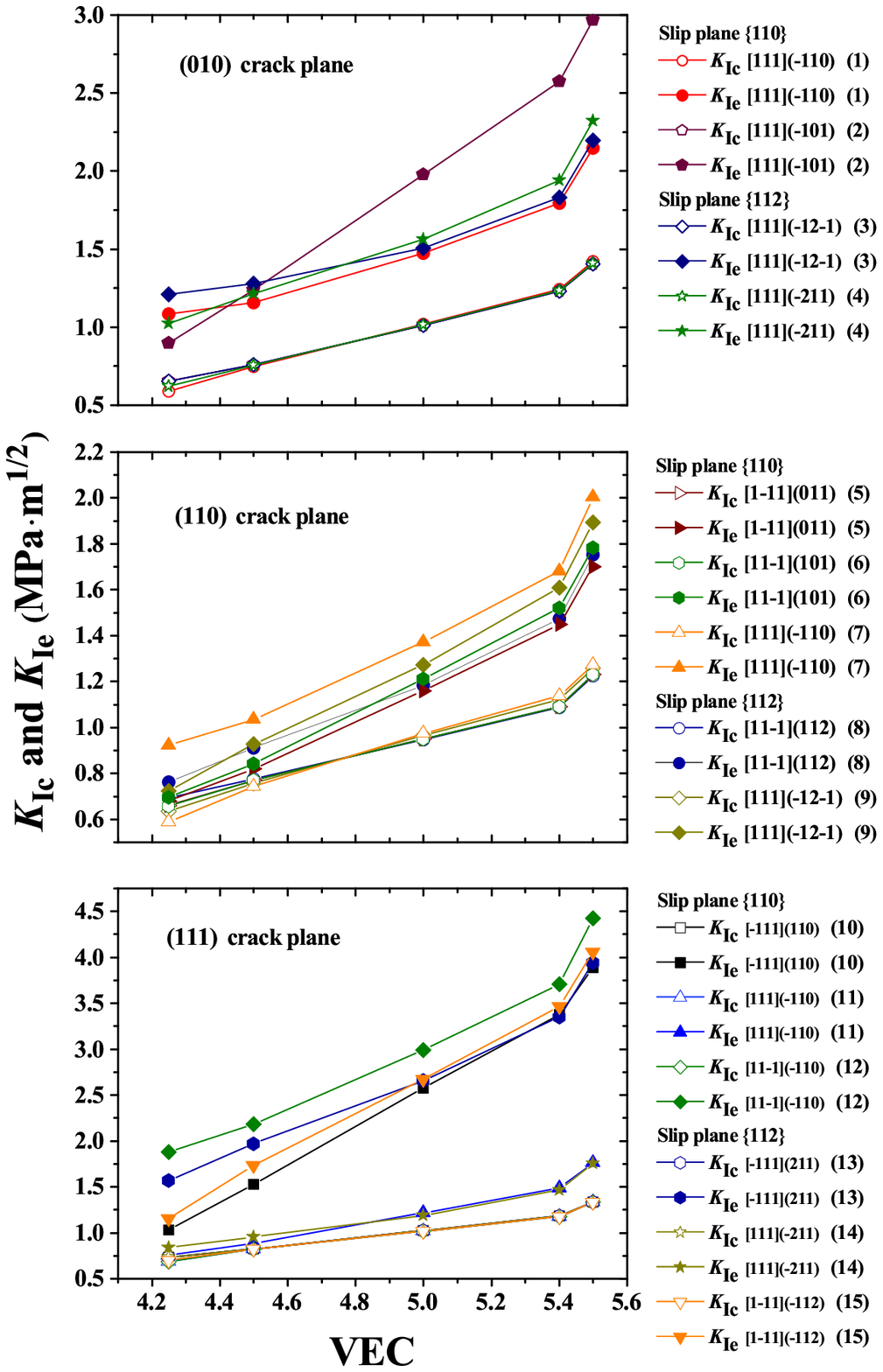}}
		\caption{\label{fig:intensity factor}The variations of the critical stress intensity factor for Griffith cleavage $K_{\RN{1}\text{c}}$ (open symbols) and  dislocation emission $K_{\RN{1}\text{e}}$ (solid symbols) with VEC of the bcc HEAs for the considered crack configurations and slip systems. Labeling is according to Table~\ref{table:orientation} with the factor $\nfrac{1}{2}$ omitted and the crack model given in parentheses. Lines guide the eye.}
	\end{center}
\end{figure}

\subsection{\label{sec:K}Critical stress intensity factors}

Using these material parameters, we derived the critical stress intensity factors for Griffith cleavage and dislocation emission in mode $\RN{1}$ loading for the considered crack configurations and slip systems (Table~\ref{table:orientation}) based on the LEFM analysis.
Figure~\ref{fig:intensity factor} shows $K_{\RN{1}\text{c}}$ (open symbols) and $K_{\RN{1}\text{e}}$ (filled symbols) for each of the three crack planes and the multiple possible slip systems as a function of the VEC, i.e., in the order HfNbTiZr, NbTiVZr, MoNbTiV, MoNbTaVW, and MoNbTaW. 

Examining Fig.~\ref{fig:intensity factor}, both $K_{\RN{1}\text{c}}$ and $K_{\RN{1}\text{e}}$ are found to monotonically increase with the VEC for all studied crack configurations and slip systems. The slopes of $K_{\RN{1}\text{e}}$ are typically steeper than those of $K_{\RN{1}\text{c}}$.
For each crack plane and alloy, the $K_{\RN{1}\text{e}}$ differ significantly across the slip systems. 
Specifically, for the cracks on the $(010)$ plane,  the $K_{\RN{1}\text{e}}$ of model 2 has a distinctly larger slope than those of models 1, 3, and 4. The main reason behind is the effect elastic anisotropy on $F_{12}$, which is further surveyed in Sec.~\ref{sec:MatAniso}. 
The slip system with the lowest $K_{\RN{1}\text{e}}$ shifts from model 2 for HfNbTiZr to model 1 for the remaining alloys. 
For the cracks on the $(110)$ plane, the five $K_{\RN{1}\text{e}}$ curves cluster and run mostly parallel, and model 5 with dislocation emission on $[\bar{1}11](011)$ has the lowest $K_{\RN{1}\text{e}}$ among all alloys.
Finally, for the $(111)$ oriented cracks, the $K_{\RN{1}\text{e}}$ fall into two groups with starkly different slopes: the increases of $K_{\RN{1}\text{e}}$ with the VEC of models 11 and 14 are much smaller than those of the other four slip systems. This leads to distinct $K_{\RN{1}\text{e}}$ values (by a factor of 2-3) for MoNbTaW.
An analysis of the parameters entering the condition for dislocation emission [Eq.~\eqref{eq:SBfinalKI}] revealed that the crack orientation is mainly responsible: models 11 and 14 are configurations, where $\theta$ is large and $\phi$ is small (thus maximizing the resolved shear stress along the slip direction), whereas for the other slip systems the smaller $\theta$ and the larger $\phi$ inevitable increase $K_{\RN{1}\text{e}}$.

In comparison, $K_{\RN{1}\text{c}}$ for Griffith cleavage under pure mode \RN{1} loading generally does not vary strongly across the various crack-tip directions for each alloy and crack plane.
These variations nevertheless reflect the fact that the Irwin energy release rate may differ for different crack extension directions (even on the same crack plane) in elastically anisotropic materials. Section~\ref{sec:MatAniso} discusses this effect of material anisotropy in more detail.

Overall, 
the trends of both $K_{\RN{1}\text{c}}$ and $K_{\RN{1}\text{e}}$ obviously correlate well with those set by the surface energies and USF energies as a function of VEC; cf.~Fig.~\ref{fig:energy}.
That is, the elastic constants, or more precisely the variables representing the elastic field solution in Eq.~\eqref{eq:SBfinalKI}, do not override these behaviors, but are responsible for the variations in $K_{\RN{1}\text{e}}$ and $K_{\RN{1}\text{c}}$ observed across the slip systems for each material.

\subsection{\label{sec:response}Ductile versus brittle crack-tip response} 

\begin{figure}[tb]
	\begin{center}
		\resizebox{!}{.5\textheight}{\includegraphics[clip]{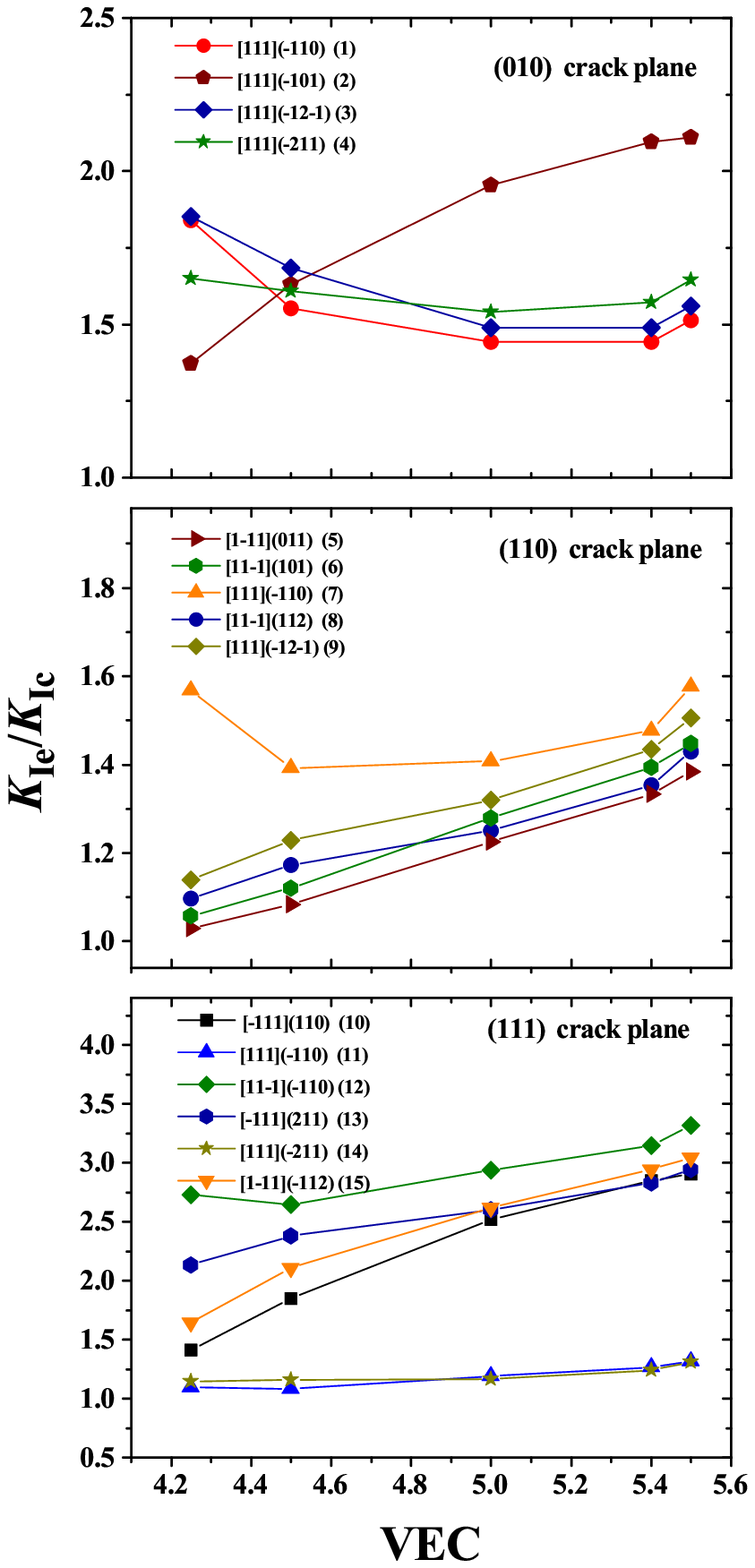}}\\
		\caption{\label{fig:ductility parameter}The variations of the stress intensity factor ratio $K_{\RN{1}\text{e}}$/$K_{\RN{1}\text{c}}$ with VEC of the bcc HEAs for the considered crack configurations and slip systems. Labeling is according to Table~\ref{table:orientation} with the factor $\nfrac{1}{2}$ omitted and the crack model given in parentheses. Lines guide the eye.}
	\end{center}
\end{figure}

The results of the previous section allow investigating the competition between the ease to plastic flow versus the ease to brittle cleavage fracture at the tip of a sharp crack subject to mode \RN{1} loading, which is embodied in the ratio of $K_{\RN{1}\text{e}}$ to $K_{\RN{1}\text{c}}$.
A $K_{\RN{1}\text{e}}$/$K_{\RN{1}\text{c}}$ ratio larger than 1 indicates that brittle cleavage is favorable, whereas $K_{\RN{1}\text{e}}$/$K_{\RN{1}\text{c}} < 1$ favors ductile response by dislocation emission.
Figure~\ref{fig:ductility parameter} shows the
$K_{\RN{1}\text{e}}/K_{\RN{1}\text{c}}$ ratio for each of the three crack planes and the multiple possible slip systems as a function of the VEC for the HEAs.

Examining Fig.~\ref{fig:ductility parameter},
for the cracks on the (010) plane, $K_{\RN{1}\text{e}}/K_{\RN{1}\text{c}}$ is clearly larger than 1 for all considered slip systems. This suggests  cleavage decohesion by crack propagation on the $(010)$ plane occurs without appreciable dislocation activity at the crack tip. 
The compositional trend of model 2 is obviously different from the other models mainly due to $K_{\RN{1}\text{e}}$ as discussed above.
For the (110) oriented cracks, the $K_{\RN{1}\text{e}}/K_{\RN{1}\text{c}}$ ratios of MoNbTiV, MoNbTaVW, and MoNbTaW are substantially larger than 1 for all five orientations, and Griffith cleavage is predicted. 
In contrast, for NbTiVZr and models 5 and 6, and for HfNbTiZr and models 6 and 8, $K_{\RN{1}\text{e}}/K_{\RN{1}\text{c}}$ approaches 1. 
In particular, $K_{\RN{1}\text{e}}/K_{\RN{1}\text{c}}$ is only marginally larger than 1 (1.02) in the case of HfNbTiZr and model 5. These are cases where the competition between dislocation emission and crack propagation is close and definite predictions on the crack tip response are difficult.

Finally, for the cracks on the (111) plane, models 11 and 14  exhibit a $K_{\RN{1}\text{e}}/K_{\RN{1}\text{c}}$ ratio slightly larger than 1 and nearly constant across all five HEAs. These two slip systems are thus predicted to marginally favor Griffith cleavage. 
$K_{\RN{1}\text{e}}/K_{\RN{1}\text{c}}$ is clearly larger than 1 for the other four slip systems and all considered alloys, and Griffith cleavage over dislocation emission is predicted.

The above predictions based on the theory of Rice and the LEFM analysis suggests 
that
these five refractory HEAs are rather intrinsically brittle at 0\,K, 
particularly those alloys with VEC$\ge$5.0, as Griffith cleavage is predicted 
for all considered orientations. Nevertheless, there are a few slip systems and 
crack configurations with $K_{\RN{1}\text{e}} /  K_{\RN{1}\text{c}} \gtrapprox 1$ 
as detailed above. 
Although the Rice theory was shown to agree fairly well with 
results of atomistic simulations~\cite{Hai:2003,Wu:2015}, definite answers on the crack 
tip response for borderline cases with $K_{\RN{1}\text{e}} \approx  
K_{\RN{1}\text{c}}$ may be difficult due to approximations of the Rice theory, 
assumptions on linear elasticity, etc. 
These cases may be resolved by means of atomistic simulations, but obviously require an accurate interatomic potential for multi-component HEAs.
The decrease of $K_{\RN{1}\text{e}} /  K_{\RN{1}\text{c}}$ with decreasing VEC determined for many models
suggests that ductilization  
may be possible for some cases, at least for cracks on (110), for HEAs in bcc phase possessing a VEC$< $4.25.

\subsection{Comparison to fracture experiments on single crystals\label{sec:comp}}

Using in situ micro-cantilever bending tests, Zou \emph{et al.}~\cite{Zou:2017} recently investigated the fracture properties of micro-meter sized, single-crystal MoNbTaW specimens containing a slit-like notch cut along a \{100\} crystallographic plane. 
A limited amount of plasticity at the crack tip before fracture was observed (river markings were seen) and a quasi-cleavage fracture mode was concluded. As brittle cleavage was suggested to be the main relevant energy consumption mechanism~\cite{Zou:2017}, the Griffith value may be a close estimate for the critical energy release rate, 
and we may compare the experimentally derived plane-strain stress intensity factor with our results.
The notch-crystal orientation in the experiments, $[\bar{1}\bar{1}0]$-$[001]$-$[\bar{1}10]$ in terms of $x_1$-$x_2$-$x_3$, is equivalent to that of model 3. 
We found a theoretical $K_{\RN{1}\text{c}}$ of 1.41\,MPa$\cdot$m$^{1/2}$ for the latter model in agreement with the experimental value in the range of 1.3-2.1\,MPa$\cdot$m$^{1/2}$ (average of 1.6\,MPa$\cdot$m$^{1/2}$ over four tests)~\cite{Zou:2017}.
We recall that $K_{\RN{1}\text{e}} > K_{\RN{1}\text{c}}$ for model 3 and MoNbTaW, and a brittle response is expected theoretically.

Several additional factors should be kept in mind when comparing theoretical and experimental stress intensity factors, and specifically the $K_{\RN{1}\text{c}}$:
\newline
(a) Our continuum analysis assumed a sharp crack tip, whereas physically realistic crack-tip geometries are likely blunt at the atomic scale. 
We note in passing that the pre-cracked specimens in the experiments of Zou \emph{et al.} had a crack-tip radius of curvature of $\approx 10$\,nm~\cite{Zou:2017}. 
The LEFM analysis may thus be extended to more realistic crack-tip geometries in order to investigate the influence of crack blunting on atomic decohesion (and dislocation nucleation). 
A modification of the present crack model seems feasible for crack-tip geometries where closed form solutions for the crack-tip stress fields exist, e.g., elliptical cracks to represent a blunted crack-tip shape. 
For the latter, Beltz \emph{et al.}'s continuum analysis for isotropic elastic media suggested that the energy release rate for crack advance is somewhat larger than the Griffith value of $2\gamma_{s}$ at above a few Burgers vectors crack-tip radius consistent with findings from atomistic simulations~\cite{Beltz:1999} (the energy release rate for dislocation emission was also reported to increase with crack-tip radius). 
\newline
(b) The theoretical predictions did not consider the effect of finite temperature on material property values, which would influence $K_{\RN{1}\text{c}}$ via temperature-dependent elastic constants and surface energies. The variations of the elastic constants and surface energies from 0\,K to ambient temperature are expected to be weak, however.\footnote{To our best knowledge, systematic measurements of elastic constants below ambient temperature for refractory HEAs have not been reported yet. Resorting to the nonmagnetic bcc refractory elements, the single-crystal elastic constants typically reduce by 2-7\,\% from 4.2 to 300\,K~\cite{Featherston:1963,Bolef:1971}. The experimental determination of surface energies is generally difficult and hampered by experimental difficulties; for metals, a widely accepted semi-empirical estimate of the surface energy decrease between 0\,K and the melting point is 10-15\,\%~\cite{Murr:1975,Tyson:1977}.}
To give a simple, but reasonable, estimate of this effect, we may assume a simultaneous decrease of $\gamma_\text{s}$ and all elastic constants by 5\,\%. This reduces the 0\,K value of $K_{\RN{1}\text{c}}$ by approximately the same relative amount to 1.34\,MPa$\cdot$m$^{1/2}$.
The rather weak temperature behavior of the Griffith level is in contrast to the effect of temperature on dislocation nucleation from a crack tip, which is a thermally activated event with expectedly stronger temperature-dependent USF energy and activation energy barrier~\cite{Rice:1994,Warner:2009}. 
Although estimating the influence of temperature on $K_{\RN{1}\text{e}}$ is beyond the scope of this work, we note that the experimentally observed quasi-cleavage fracture of MoNbTaW would be consistent with the fact that $K_{\RN{1}\text{e}} > K_{\RN{1}\text{c}}$ at ambient temperature. That is, the critical load level for \emph{spontaneous} dislocation emission is above that of cleavage, and thermally activated nucleation, which may already take place at lower load levels, did not occur within the finite waiting time of the experiment (determined by the loading rate).
\newline
(c) Loading conditions and crack configurations in the experiments may not be as ideal as assumed in the theoretical analysis. Specifically, loading may deviate from a pure mode \RN{1} tensile opening and a mixed-mode loading may occur instead. The latter contains shear components in addition, the effect of which on the critical stress intensity factors is further analyzed theoretically in the next section.
Some degree of misalignment between the crack-tip direction and the crystallographic crack plane may also occur, which was suggested in Ref.~\cite{Zou:2017} to be responsible for the observed river markings.
\newline
(d) Our theoretical analysis relies on a quasi-static picture, wherein 
dislocation emission and crack extension are mutually exclusive processes. 
There exist materials and conditions of temperature and load rate, where dislocation emission and crack extension need not be mutually exclusive~\cite{Jokl:1980b}. This leads to a dynamical scenario the theoretical modeling of which would, for instance, involve a dynamical treatment of dislocation motion and a differentiation between pre-existing and injected cracks~\cite{Jokl:1989}.
Furthermore, our theoretical analysis assumes a pristine solid free of lattice dislocations, whose stress field may shield the critical stress intensity factors. 
In the experiments of Zou \emph{et al.}~\cite{Zou:2017}, the bulk MoNbTaW sample was long-term annealed prior to micro-cantilever fabrication and mechanical testing, and we may assume that the specimens were loaded under dislocation-low (if not dislocation-free) conditions.

We note in passing that when brittle cleavage is the main relevant energy consumption mechanism, as supposedly in MoNbTaW, the Griffith level may provide an experimental value of the surface energy of the crack plane. Assuming the experimental $K_{\RN{1}\text{c}}$ of 1.6\,MPa$\cdot$m$^{1/2}$, Eq.~\eqref{eq:help1} gives a surface energy of 3.9\,J/m$^2$ for the $(010)$ facet of MoNbTaW.

\subsection{Deviation from pure mode \RN{1} loading\label{sec:deviation}}

It is instructive to briefly survey the sensitivity of $K_{\RN{1}\text{e}} / K_{\RN{1}\text{c}}$ to deviations from pure mode \RN{1} loading, which may, for instance,  unintentionally occur due to misorientation in cantilever experiments (it should be noted that the critical stress intensity factors are denoted as before despite mixed-mode loadings).                                                                                                 
To this end, we introduce fractional shear loadings as in Ref.~\cite{Rice:1992}, $K_\RN{2} = xK_\RN{1}$ and $K_\RN{3} = zK_\RN{1}$. 
Letting ${\bm K} = K_\RN{1} [x,1,z]^T \equiv K_\RN{1} {\bm k}$, Eq.~\eqref{eq:help1} becomes                                                                         
\begin{align}                                                                                      
 K_{\RN{1}\text{c}} &= \sqrt{ \frac{ 2\gamma_{\text{s}}}{ {\bm k}^T {\bm \Lambda} {\bm k}  } }.            
 \label{eq:KIcmixed}
\end{align}   
Similarly, if dislocation emission is interpreted in terms of the effective shear stress intensity factors, 
\begin{align}                                                                                         
  K_{\RN{1}\text{e}} &= \frac{\sqrt{\gamma_{\text{usf}}\, {\bm s}^T(\phi) {\bm \Lambda}^{(\theta)-1} {\bm s}(\phi) }}{ {\bm s}^T(\phi) {\bm F} {\bm k} }.
  \label{eq:KIemixed}
\end{align} 
As before, the $F_{\alpha\beta}$ functions are those for pure mode $K_\beta$ loading on the main crack.

\begin{table}[thb]
\caption{\label{table:deviation}Ratio $K_{\RN{1}\text{e}} / K_{\RN{1}\text{c}}$ [from Eqs.~\eqref{eq:KIcmixed} and~\eqref{eq:KIemixed}] for deviation from pure mode \RN{1} loading for two HEAs and two crack configurations (${\bm K} = K_\RN{1} {\bm k}$).}
 \begin{tabular}{ccccc}
 \toprule
           & \multicolumn{2}{c}{HfNbTiZr} & \multicolumn{2}{c}{MoNbTaVW} \\
           \cmidrule(lr){2-3}\cmidrule(lr){4-5}
 ${\bm k}^T$ & Model 5 & Model 11 & Model 5 & Model 11\\
 \midrule
 $[0.00,1,0]$ & 1.03 & 1.15& 1.33 & 1.26\\
 $[0.05,1,0]$ & 1.00 & 1.22& 1.30 & 1.33\\
 $[0.10,1,0]$ & 0.97 & 1.29& 1.26 & 1.41\\
 \bottomrule
 \end{tabular}
\end{table}

In order to illustrate the effect of deviations from pure mode \RN{1} loading,
Table~\ref{table:deviation} compares $K_{\RN{1}\text{e}} / K_{\RN{1}\text{c}}$ when the in-plane shear $K_{\RN{2}}$ is 5\,\% or 10\,\% of $K_{\RN{1}}$ (i.e., $x=0.05$ or $x=0.1$) with the results of pure mode $\RN{1}$ loading (i.e., $x=0.0$) for two of the current HEAs and crack models 5 and 11. 
It should be noted that the norm of vector ${\bm k}$ cancels out in taking the ratio $K_{\RN{1}\text{e}} / K_{\RN{1}\text{c}}$. For the sake of completeness, the respective $K_{\RN{1}\text{e}}$ and $K_{\RN{1}\text{c}}$ values are listed in Table~\ref{table:deviationdetail}.

It follows that $K_{\RN{1}\text{e}} / K_{\RN{1}\text{c}}$ is sensitive to deviations from pure mode \RN{1} loading, in particular so for model 11, whereas the absolute variations with $x$ for HfNbTiZr and MoNbTaVW are of comparable magnitude.
Analysis of the respective $K_{\RN{1}\text{e}}$ and $K_{\RN{1}\text{c}}$ values (Table~\ref{table:deviationdetail}) shows that finite in-plane shear mostly perturbs $K_{\RN{1}\text{e}}$.
Furthermore, the critical ratio may decrease, in the case of HfNbTiZr and model 5 below the ratio of 1, or increase (e.g., model 11) with finite in-plane shear.
This behavior originates from $F_{11}$ (i.e., the $\theta$-dependence of $\sigma_{\theta r}$ for $K_\RN{2}$), which takes positive values below a critical angle and negative ones above (at the critical angle, $F_{11}=0$ and the finite in-plane shear would have no effect on the nucleation condition). 
In the formulation for isotropic elastic media, $F^\text{iso}_{11}$ equals $\cos (\theta /2 ) [1-3\sin^2(\theta/2)]$ and changes sign at approximately 70.5$^\circ$. 
For the present materials, $F_{11}$ is a monotonically decreasing function in $\theta \in [0,90^\circ]$ and the mixed-mode nucleation deviates most strongly from the pure mode $\RN{1}$ results when $\theta = 0$ or $90^\circ$.

\begin{figure*}[htb]
	\begin{center}
		\begin{tabular}{@{}lll@{}}
			(a) & (b)& (c) \\
			\resizebox{!}{0.15\textheight}{\includegraphics[clip]{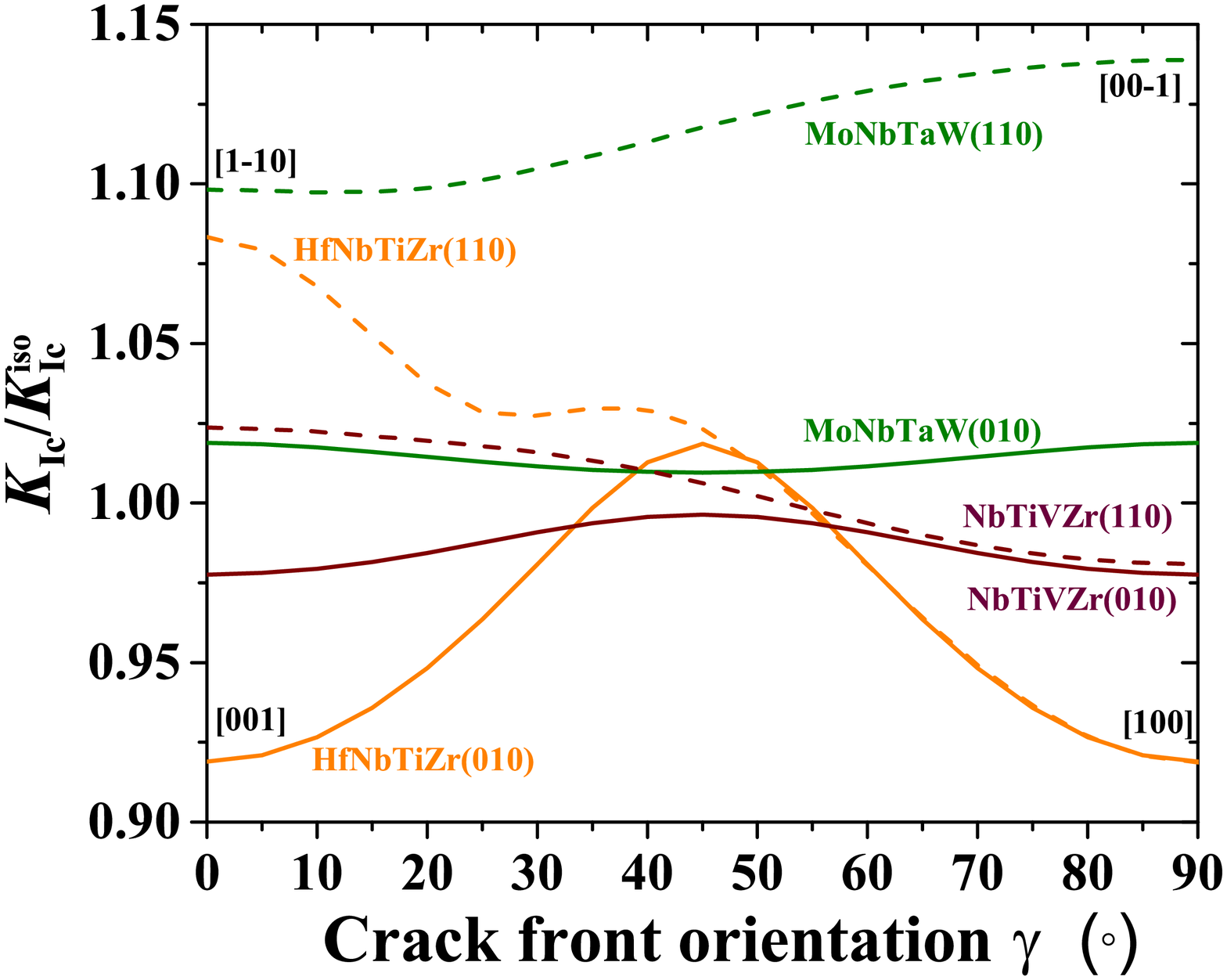}}&
			\resizebox{!}{0.15\textheight}{\includegraphics[clip]{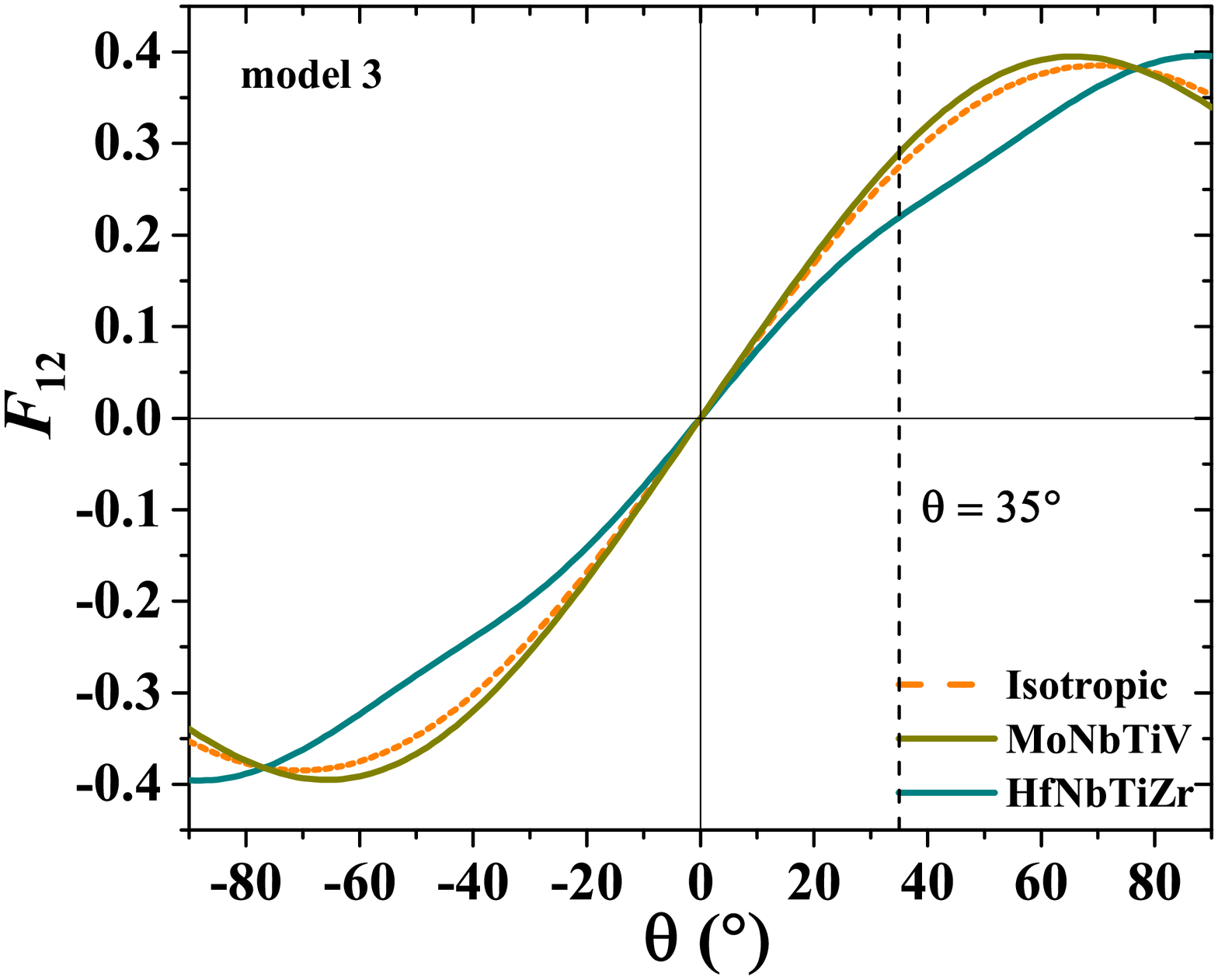}}&
			\resizebox{!}{0.15\textheight}{\includegraphics[clip]{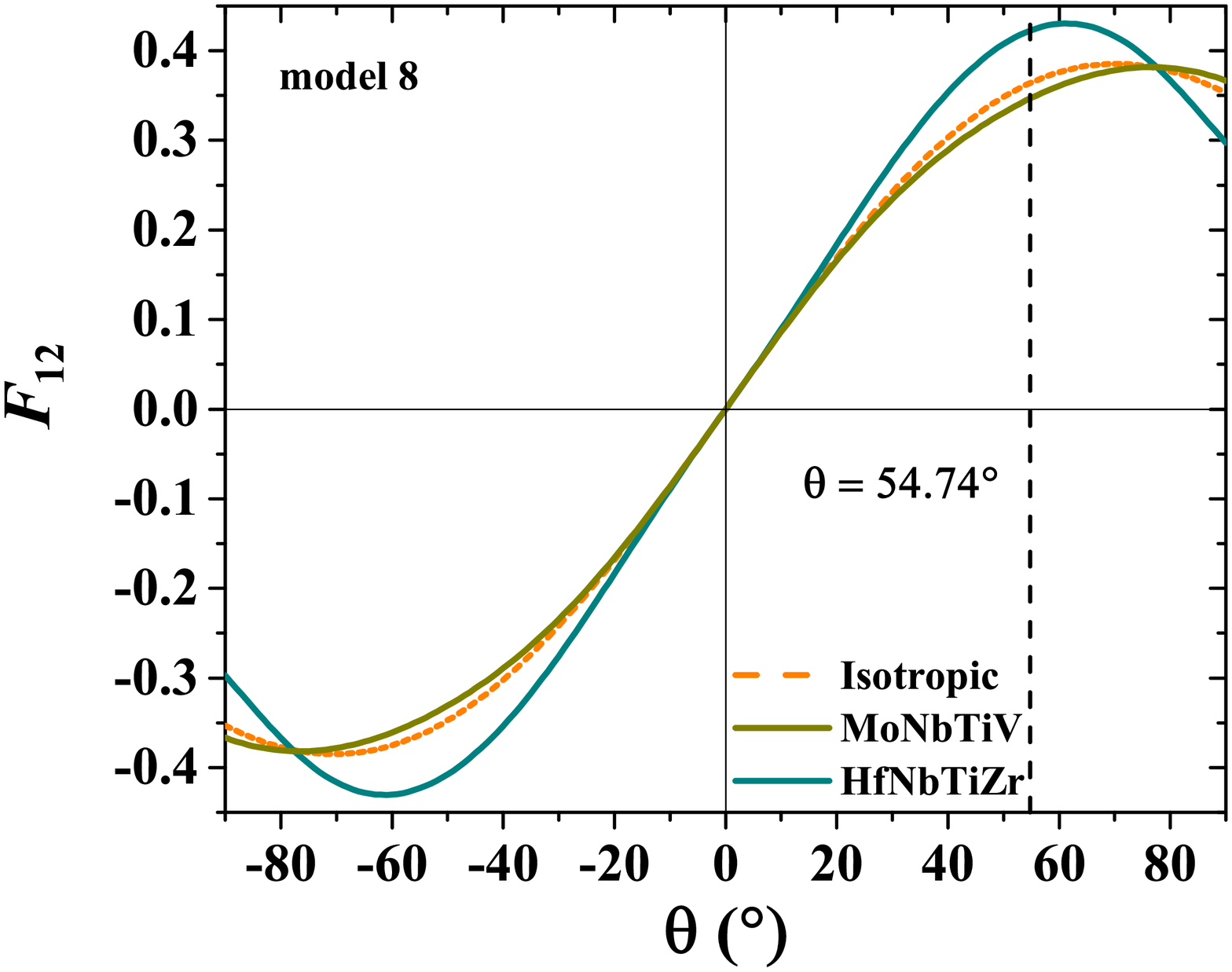}}\\			
		\end{tabular}
		\caption{\label{fig:aniso}Effect of material anisotropy in selected refractory HEAs: (a) Variation of critical stress intensity factor for crack extension with crack front direction $K_{\RN{1}\text{c}}(\gamma)/K^\text{iso}_{\RN{1}\text{c}}$. $\gamma$ measures the angle between high-symmetry directions ($[001]$ and $[1\bar{1}0]$ for the $(010)$ plane and the $(110)$ plane, respectively) and $x_3$ rotated counter-clockwise around the crack plane normal ($x_2$). (b-c) $\theta$ dependence of the normalized, singular shear stress $\sigma_{\theta r}$ in pure mode $\RN{1}$ loading [$F_{12}(\theta)$] for crack orientation models 3 and 8. The vertical dashed line denotes the angle between the crack plane and the actual slip plane for both models. The isotropic elastic solution [$\sin (\theta / 2) \cos^2 (\theta / 2 ) $] is shown for comparison. }
	\end{center}
\end{figure*}

\subsection{\label{sec:MatAniso}Effect of material anisotropy}

The anisotropic effect is surveyed in two parts by comparing to results of the formulation for isotropic elastic media. 

The Griffith energy release rate may change with crack plane, but it is independent on crack front orientation for a specific crack plane. 
The Irwin energy release due to elastic forces on the crack [Eq.~\eqref{eq:defG}] may, however, depend on crack front orientation (through ${\bm \Lambda}$). 
Thus, there may exist preferred crack front orientations for cleavage fracture in anisotropic materials.

In order to determine the anisotropy of $K_{\RN{1}\text{c}}$ and the preferred crack front orientations for some of the current HEAs, we plot the variation in critical stress intensity factor for crack extension with crack front orientation $K_{\RN{1}\text{c}}/K^\text{iso}_{\RN{1}\text{c}}$ in Fig.~\ref{fig:aniso}(a), where $K^\text{iso}_{\RN{1}c}$ is the intensity factor computed from isotropic elastic moduli. $K^\text{iso}_{\RN{1}\text{c}}$ was obtained by equating the energy release rate for isotropic, linear elastic media $G = (K^\text{iso}_{\RN{1}})^2(1-\nu^2)/E$ with Eq.~\eqref{eq:GriffithG}, where $E$ and $\nu$ are Young's modulus and Poisson's ratio, respectively, determined from the computed single-crystal elastic constants through Reuss-Voigt-Hill averaging~\cite{Hill:1952}.

It follows from Fig.~\ref{fig:aniso}(a) that the preferred crack front orientation, i.e., the minimum of $K_{\RN{1}\text{c}}/K^\text{iso}_{\RN{1}c}$, is alloy specific.
For cleavage fracture on the $(010)$ plane in HfNbTiZr and NbTiVZr, for which $A_{\text{Z}}>1$, the preferred direction is parallel to $[001]$, whereas the preferred direction is parallel to $[101]$ in MoNbTaW ($A_{\text{Z}}<1$).
The directions parallel to $[001]$ are preferred for cleavage on $(110)$ in HfNbTiZr and NbTiVZr, whereas the minimum of $K_{\RN{1}\text{c}}/K^\text{iso}_{\RN{1}c}$ in MoNbTaW does not lie parallel to a high-symmetry direction (minimum at approximately 12$^\circ$ measured between $[1\bar{1}0]$ and $x_3$ rotated counter-clockwise around $x_2$).
HfNbTiZr exhibits the largest variations in $K_{\RN{1}\text{c}}$ across the plotted crack extension directions, consistent with the fact that it is the least elastically isotropic ($A_\text{Z}=4.3$).

Similarly, Rice's nucleation condition [Eq.~\eqref{eq:Rice}] depends on the crystallographic slip system, but 
it is not affected by the relative orientation of the slip path to the crack front direction and the load applied to it, since dilational displacements did not enter into the formulation of the mechanics problem directly.
Apart from ${\bm \Lambda}$, elastic anisotropy nonetheless affects the angular dependence of the singular shear stress $\sigma_{\theta r}$ acting in the slip plane, i.e., $F_{12}(\theta)$ in previous notation.
This anisotropic effect is illustrated for crack models 3 and 8 in Fig.~\ref{fig:aniso}(b) and (c), respectively.
It follows that the anisotropic result for HfNbTiZr has significantly different shapes from the curve obtained using isotropic elasticity, whereas the anisotropic result for MoNbTiV ($A_{\text{Z}}=0.68$) only slightly differs from the isotropic formulation. As clearly illustrated for HfNbTiZr, the isotropic result can overestimate or underestimate the anisotropic curves, and the $\theta$ that maximizes $F_{12}$ can significantly differ in the anisotropic and isotropic formulations.

\subsection{\label{sec:lee}Local environment effect}

Fluctuations in local alloy composition on small length scales (local chemical environment) may translate into spatial fluctuations of dislocation properties on a similar length scale, e.g., variation in the width of the stacking fault ribbon in a face-centered cubic HEA~\cite{Smith:2016}.
Nucleation of dislocation embryos from 
crack tips is a localized process and likely susceptible to local fluctuations in alloy composition on the slip plane.
That is, an incipient dislocation configuration in an HEA might favorably nucleate in regions, where fluctuations in local alloy composition locally reduce resistance to dislocation emission. 
Similarly, resistance to dislocation emission might be greater than average in other places along the crack tip. 

In what follows, we first seek to state more precisely what is understood under localized nucleation and how it is related to variation of USF energy.
We then investigate the effect of fluctuations in local alloy composition on the competition between crack advance and nucleation of two-dimensional dislocation geometries from a straight crack tip. To this end, we analyze the results of   additional calculations performed with super cells (see below).

It is helpful for comprehending the problem to glance at a three-dimensional dislocation slip instability first.
Results from numerical and atomistic investigations, carried out for elements hitherto, identified three-dimensional incipient dislocation configurations as dislocation loops, and three-dimensional saddle point configurations as localized outward protrusions from the straight crack tip~\cite{Rice:1994,Xu:1997,Zhu:2004}.
Lateral and forward extensions of the critical saddle point state 
were shown to vary with material, crystal structure, and loading level. 
It will be useful to specify its approximate size in the athermal limit.
The extensions of the saddle point state (the part made up of core material) were determined for few bcc metals in Refs.~\cite{Rice:1994,Xu:1997}. Near the critical loading, the lateral extension ranged approximately from 10 to 17\,$b$ and the forward extension to approximately 3-5\,$b$, respectively ($b$ is the magnitude of the Burgers vector). 
As the three-dimensional embryos seen approximate to somewhat a half-ellipse~\cite{Rice:1994,Xu:1997,Zhu:2004}, the saddle point configuration has the approximate size of $(90/4$-$255/4)\pi a^2$ ($a$ denotes the bcc lattice parameter).
In the case of concentrated alloys, we expect that the extensions of the saddle point state and its nucleation barrier
are also influenced by the local chemical configuration on the slip plane.
For the sake of the further discussion, we tentatively 
accept above estimates as approximate, though representative, for a typical bcc metal, including the HEAs considered here. 
(The results and discussion which follow do not critically depend on these values.)

A two-dimensional dislocation slip instability results in the nucleation of a straight line dislocation parallel to the crack tip (homogeneous nucleation). 
This implies that an incipient dislocation configuration in a random solid solution experiences a slip potential that is configuration-averaged over the dislocation length considered, 
 motivates to explore its local-environment dependence. 
In contrast to the three-dimensional case, where the saddle point state is a localized protrusion, 
the freedom in choosing the length of a straight dislocation instability motivates studies of the local-environment dependence of the nucleation instability condition on dislocation length.
A sensible dislocation line length for the two-dimensional case may be based on results of the three-dimensional analysis~\cite{Rice:1994}, i.e., the lateral extension of the saddle point state, 
allowing direct comparison between results of two-dimensional and three-dimensional analysis.

The USF energy measures the resistance to dislocation nucleation in the Rice theory~\cite{Rice:1992}, and its value may depend on the chemical configuration in the vicinity of the fault plane.
As variation of planar fault energies in concentrated alloys is related to fault area considered, the freedom in choosing its value motivates studies of the local-environment dependence of the USF energy on fault area.
For the above reason, 
a sensible fault area for the two-dimensional slip instability
may similarly be based on results of the three-dimensional analysis, i.e., the size of the three-dimensional dislocation embryo as discussed above.
Lastly, fully configuration-averaged material parameters 
represent a well-defined limit, i.e., averages over infinitely extended (practically very large) fault areas.
Thus, it is appropriate to define the average dislocation nucleation resistance for a slip system based on the configuration-averaged USF energy of that slip system.

It should be noted that mapping USF energies with all-electron total energy methods is currently not feasible for areas larger than $\approx 15$-$30a^2$. 
Our results presented next allow to estimate the effect of fluctuations in local alloy composition on fault area in this range. 
These areas are smaller than the size of the saddle point configuration of an embryonic dislocation loop estimated above.
One expects larger variation of the USF energy for smaller fault area considered (self-averaging), thus a possibly larger effect of such fluctuations on the crack-tip competition.

In order to determine these effects on the competition between crack advance and dislocation nucleation, 
the local-environment dependencies of the surface energies for the \{100\} and \{110\} terminations, and the USF energies for the $\langle 111\rangle\{110\}$ and $\langle 111\rangle\{112\}$ slip systems were investigated for HfNbTiZr.
The details of these calculations are summarized in~\ref{sec:appLF}, where we also present the distributions of surface energies and USF energies obtained, the distribution averages, standard deviations, minimum and maximum values, and a comparison to the CPA results. Note that these CPA results were obtained with the same computational setup as the super cell calculations.
We decided on HfNbTiZr since this alloy shows a few borderline cases ($K_{\RN{1}\text{e}}\approx K_{\RN{1}\text{c}}$; see Sec.~\ref{sec:response}), where the expected local environment effect could lead to a qualitative change of the predicted failure behavior. 
Furthermore, crack advance was assumed to be self similar, i.e., all parts of the region ahead of the crack tip cleave homogeneously. Thus, a configuration-averaged surface energy (CPA result or distribution average) of the crack plane remains the relevant material parameter for crack extension. 
Exploring the local-environment dependence of the \{111\} surface energy required larger super cells than was computationally feasible and was not carried out.
In order to include \{111\} cleavage planes in the analysis nevertheless, the CPA surface energies were employed, where super cell results were not attainable (concerns models 10-15); see~\ref{sec:appLF} for additional comments and justification.
The elastic constants employed in the LEFM analysis were those determined through the CPA (Table~\ref{table:lattice_constants}). 
To summarize, the planar fault energies employed in the analysis below are those listed in~\ref{sec:appLF} (\{111\} surface energies in Table~\ref{table:lattice_constants}).

Table~\ref{table:fluc} shows three sets of $K_{\RN{1}\text{e}}/K_{\RN{1}\text{c}}$ ratios for each of the 15 crack configuration and slip systems considered. 
Therein, we first contrast the predicted ratios using the CPA results for the planar fault energies with those obtained from super cell distribution averages.
Relative to the CPA results, using distribution averages results in slightly smaller ratios throughout. The decreases are in the range of 0.02-0.07 over the 15 models, and the largest relative change amounts to -3\,\%.
Notably, $K_{\RN{1}\text{e}}/K_{\RN{1}\text{c}}$ decreases to slightly below unity for model 5, but remains a borderline case.
Thus, attaining configuration-averaged materials parameters through the CPA or super cell distributions yields rather similar $K_{\RN{1}\text{e}}/K_{\RN{1}\text{c}}$ ratios for the crack-tip competition on the average.

To mimic a locally reduced resistance to dislocation emission, we now consider $K_{\RN{1}\text{e}}/K_{\RN{1}\text{c}}$ obtained from the minimum USF energy of each distribution; see Table~\ref{table:fluc}. 
Using the minimum USF energy expectedly lowers the critical ratios of stress intensity factors relative to those characterizing the average crack-tip competition, i.e., the shifts are towards a greater propensity for dislocation emission.
With respect to super cell distribution averages, the decreases are slight and in the range of 0.02-0.10 over the 15 models. The most significant relative change amounts to -4\,\%.

These results indicate that the variation in local alloy composition on the crack-tip competition in HfNbTiZr are of minor importance on the absolute scale of the $K_{\RN{1}\text{e}}/K_{\RN{1}\text{c}}$ ratios predicted (i.e., 
those derived from configuration-averaged material parameters).
The overall brittle crack-tip behavior in HfNbTiZr is not changed.
In the case of close crack-tip competition ($K_{\RN{1}\text{e}}/K_{\RN{1}\text{c}}\approx 1$), these fluctuations may locally reduce resistance to dislocation emission just enough to render a ductile response slightly more favorable than brittle cleavage. Definite answers on the crack-tip behavior in such cases are still difficult, since the decreases found are small. 
We expect that the variation of USF energy decreases for larger fault areas, 
which in turn diminishes the influence of local fluctuation in alloy composition on the nucleation barrier.
For the other alloys considered here, we expect that fluctuations in local alloy composition have a similarly small effect on the crack-tip competition and do not change the dominant crack-tip behavior.

\begin{table*}[tbh]
        \caption{\label{table:fluc}Local environment effect on the ratios of $K_{\RN{1}\text{e}}$ to $K_{\RN{1}\text{c}}$ for HfNbTiZr and models 1-15. Contrasted are results obtained from CPA planar fault energies, super cell distribution averages (denoted by aver), and the distribution minimum of the USF energies (denoted by min).}
        \begin{tabular}{ll*{4}{c}|*{5}{c}|*{6}{c}}
         \toprule
         \multicolumn{2}{c}{Data} & \multicolumn{15}{c}{$K_{\RN{1}\text{e}}/K_{\RN{1}\text{c}}$ for model} \\
         \cmidrule(lr){1-2}\cmidrule(lr){3-17}
         $\gamma_{\text{usf}}$ & $\gamma_{\text{s}}$ & 1 & 2 & 3 & 4 & 5 & 6 & 7 & 8 & 9 & 10 & 11 & 12 & 13 & 14 & 15\\
         \midrule
         CPA & CPA & 1.79 & 1.34 & 1.87 & 1.49 & 1.01 & 1.08 & 1.61 & 1.11 & 1.37 & 1.43 & 1.11 & 2.96 & 2.13 & 1.14 & 1.61\\
         aver & aver & 1.77 & 1.32 & 1.81 & 1.45 & 0.98 & 1.06 & 1.57 & 1.06 & 1.32 & 1.40 & 1.09 & 2.90 & 2.06 & 1.10 & 1.55 \\
         min & aver & 1.70 & 1.27 & 1.76 & 1.41 & 0.95 & 1.02 & 1.52 & 1.04 & 1.28 & 1.35 & 1.05 & 2.80 & 2.00 & 1.07 & 1.51 \\
         \bottomrule
        \end{tabular}
\end{table*}

\subsection{Remarks on surface ledges, crack-tip twinning, and finite-temperature effects\label{sec:remarks}}

\paragraph{Surface ledges}The Rice approach to mode $\RN{1}$ loading does not consider the effect of surface ledge formation during the dislocation emission process. This ledge energy typically amounts to a fraction of the surface energy and will affect the resistance to dislocation nucleation~\cite{Zhou:1994,Xu:1995,Juan:1996,Andric:2017}. This process can give an important contribution to the strain energy release rate in materials whose ratio of surface energy to USF energy is large. This seems to be the case for face-centered cubic metals and observed slip systems, where this ratio is typically 4-9.
In a typical bcc metal, the ratio of surface energy to USF energy is smaller, approximately 2-4. Whether surface ledge formation yields a significant contribution to the strain energy release rate in bcc metals in presently unknown and will be addressed in future work. 

\paragraph{Crack-tip twinning}
Atomistic and multiscale simulations for bcc Fe indicate that 
crack-induced nucleation of \{112\} twins competes with full dislocation slip on \{112\} planes and crack propagation (e.g.,~\cite{Machova:2004,Gordon:2007,Vatne:2011}).
Some refractory HEAs (e.g., HfNbTaTiZr~\cite{Senkov:2011}) show deformation twinning from bulk sources, and may be susceptible to twinning processes from cracks.
To theoretically address deformation twinning ahead of crack tips, and  
discern between the conditions that favor either of the two slip processes, a more general approach is required. 

Twinning in bcc metals differs from full slip in that 
the incipient microtwin involves coordinated shearing on multiple, interacting slip planes (a single-layer stacking fault on \{112\} planes is considered to be unstable)~\cite{Vitek:1970,Christian:1995}. 
In contrast to often modelled single dislocation emission, multi-plane slip models for complex deformation features are scarce.
Notably, Beltz \emph{et al.}~\cite{Beltz:2005} developed a multi-plane continuum model to handling nonlinear effects associated with extended dislocation cores and micro-twin formation in bcc crystals.
The theory originally constructed for mode $\RN{2}$ loading uses concepts from the Peierls-Nabarro framework, but unfortunately is not straightforwardly applicable to our case. The theory requires extension to mode ${\RN{1}}$ loading, twinning on inclined slip planes, and anisotropic linear elastic media. 

Modeling deformation twinning in bcc crystals also necessitates a realistic and robust constitutive response for coordinated $\langle 111\rangle\{112\}$ slip and $n$-layer stacking faults, somewhat analogous to the generalized stacking fault energy for the case of single-plane slip. From the energy potential one may obtain the displacements and interplanar shear stresses for the active number of interacting slip planes by numerical analysis~\cite{Beltz:2005}, the  first stable multi-layer stacking fault to occur on \{112\} planes, and the morphology of twinning interface (i.e., reflection or isosceles). 

We anticipate that twinning and full slip on \{112\} planes are not ``either-or'' processes. 
The crack-tip response is expected to depend on the the specific constitutive response, crack orientation (including consideration of shearing in the twinning or anti-twinning sense), and loading mode.

\paragraph{Finite-temperature effects}
The theoretical analysis presented applies to zero temperature. Intrinsic ductilization in these bcc materials may occur at finite temperature. 
An improved prediction could be achieved by determining temperature-dependent material parameters, and comparing the critical load levels for cleavage and spontaneous dislocation emission from a crack tip at that temperature.
Since the load level for cleavage is bound from below by the Griffith value, it can not decrease arbitrarily (a rough estimate at 300\,K was provided in Sec.~\ref{sec:comp}). 
It is expected that the critical load level for spontaneous nucleation decreases more rapidly with temperature than that of cleavage.

To our best knowledge, one serious attempt has been made to determine the temperature dependence of the USF energy, not for a bcc crystal but for face-centered-cubic Al and Ni through classical molecular dynamic simulations and the embedded-atom model~\cite{Warner:2009}. 
Therein, a reduction of the USF energy (corresponding to a $\langle 112\rangle/6$ displacement on \{111\} planes) by roughly 25~\% for a 
temperature increase from 0 to 300\,K was reported.
The lack of systematic investigations makes it difficult to judge whether this estimate is representative, 
but tentatively accepting the order of magnitude to be correct affirms above expectation.

For the present, the temperature dependence of the material parameters required is not computationally accessible to direct first-principles simulations. 
There are practical challenges in lattice-dynamic calculations for chemically disordered systems, as only small simulation cells are computationally feasible. 
The computational hurdle is compounded by the fact that explicit lattice anharmonicity can not safely be neglected at above approximately 0.6 homologous temperature, consideration of which requires long-time molecular dynamics simulations. 
Standard lattice dynamic calculations and free energy calculations are ruled out for the unstable stacking position. Instead, one may resort to thermodynamic integration of the internal energy in combination with long-time molecular dynamics simulations (as done in Ref.~\cite{Warner:2009}).

Thermal activation will allow dislocation nucleation at a finite rate at load levels below that of instantaneous emission, and is thus expected to lower the critical load level more rapidly~\cite{Rice:1994,Xu:1997}.
Analysis of spontaneous emission is often aimed at identifying the saddle point configuration and corresponding activation energy barrier as a function of temperature and load. The nucleation event probability is then frequently estimated from transition state theory. 
Based on the Peierls concept, Rice and Beltz~\cite{Rice:1994} estimated the saddle point configuration and associated activation energy for homogeneous nucleation in elastically isotropic materials by recourse to the USF energy. The advantage of this theory is that the activation energy barrier can be approximated with a closed-form expression employing (ideally, temperature-dependent) material parameters. Unfortunately, this analysis was developed only for co-planar slip and crack planes and is not straightforwardly applicable to our case.

Classical atomistic simulations obviously lower the computational hurdle, at the expense of accuracy, but may be applied to various solid state phenomena, including estimating
the temperature dependence of material parameters or the saddle point configuration for nucleation from crack tips, and enabling direct dislocation nucleation simulations.
However, a persistent challenge is to find accurate interatomic potentials for multi-component alloy systems including HEAs.

\section{\label{sec:Conclusions}Conclusions}

We investigated the competition between cleavage decohesion and dislocation emission from a sharp crack tip for the bcc refractory HEAs HfNbTiZr, MoNbTaVW,  
MoNbTaW, MoNbTiV, and NbTiVZr.
Dislocation nucleation, when measured by configuration averages of the nucleation condition, was found to be unfavorable in all 15 considered crack configurations and slip systems, but borderline cases ($K_{\RN{1}\text{e}} \approx  K_{\RN{1}\text{c}}$) exist in HfNbTiZr and MoNbTaVW.
Fluctuations in local alloy composition, investigated for HfNbTiZr, can locally reduce the resistance to dislocation emission relative to the configuration average, but generally had a minor effect on $K_{\RN{1}\text{e}} /  K_{\RN{1}\text{c}}$.
Moreover, such fluctuations did not change the dominant crack-tip response but for one crack orientation, where the crack-tip competition was already close.
The absence of appreciable dislocation plasticity near the crack tip thus suggests  intrinsically brittleness and low crack-tip fracture toughness in these materials. 
In the case of MoNbTaW, where an experimental $K_{\RN{1}\text{c}}$ value is available for a crack on the \{100\} plane, theoretical and experimental results agree. 

While the theoretical predictions made may be refined in future work, for instance along the lines discussed in Sec.~\ref{sec:remarks}, 
an experimental step towards verifying the predictions may be to
conduct fracture experiments, similar to those performed in Ref.~\cite{Zou:2017}, for an alloy with overall lower $K_{\RN{1}\text{e}} /  K_{\RN{1}\text{c}}$ ratios than MoNbTaW, i.e., HfNbTiZr or NbTiVZr. 
Moreover, future theoretical investigations of the kind carried out here may be broadened to non-equimolar refractory HEAs.

A challenging proposition would be to determine to 
what degree the here obtained results on the intrinsic fracture behavior surfaces in the mechanical response of bulk samples. 
It seems reassuring that the available literature on refractory HEAs (as recently reviewed in Ref.~\cite{Senkov:2018}) points to HfNbTiZr and derived compositions (e.g., HfNbTaTiZr and Hf$_{0.5}$Nb$_{0.5}$Ta$_{0.5}$Ti$_{1.5}$Zr, not investigated here) as currently being among the best room-temperature ductile refractory HEAs (tensile ductility larger than 10\,\%, compressive ductility in excess of 50\,\%~\cite{Senkov:2011b,Wu:2014,Sheikh:2016,Schuh:2018}). 
(Refractory HEAs experiencing strain-induced or stress-induced phase transformations can show higher tensile elongation at fracture.)
However, the notably better performance of HfNbTiZr and derived alloys is likely governed by complex changes in dislocation core structure and dislocation mobility, besides crack-tip phenomena.

\section{Acknowledgments}
The Swedish Research Council, the Swedish Steel Producers' Association, the Swedish Foundation for Strategic Research, the Swedish Foundation for International Cooperation in Research and Higher Education, and the Hungarian Scientific Research Fund (research project OTKA 128229) are acknowledged for financial support.
The simulations were performed on resources provided by the Swedish National Infrastructure for Computing at the National Supercomputer Centre in Link\"oping.

\appendix
\section{\label{sec:app}A crack in anisotropic linear elastic media}

This brief, self-contained summary is mainly based on the comprehensive treatise on anisotropic elasticity from Ting~\cite{Ting:1996}, as well as Bower's~\cite{Bower:2009} text book and Hwu's~\cite{Hwu:2010} exposition.

Let $u_{i}$, $\epsilon_{ij} = ( u_{i,j} + u_{j,i})/2$, $\sigma_{ij}$, and $C_{ijkl}$ be the components of the displacement field, strain tensor, stress tensor, and tensor of elastic moduli, respectively, for anisotropic elastic media in a fixed Cartesian coordinate system ($i=1,2,3)$ (cf. Fig.~\ref{fig:geometry}).

\paragraph{Basic equations}
In the LEFM analysis, one seeks the elastic displacement field ${\bm u}$ and the stress function ${\bm \Phi}$ for the governing equations of static equilibrium~\cite{Ting:1996},
\begin{align}
  C_{ijkl}u_{k,lj} &= 0, \text{ or } \sigma_{ij,j} = 0,
  \label{eq:problem} 
\end{align}
where a comma denotes differentiation (e.g., '$,i$' with respect to $x_i$) and repeated indices are summed over. 
We search for generalized plane strain deformations for which the displacement field has the form $u_i=u_i(x_1,x_2)$, and in addition  $\epsilon_{33}=0$.
The stresses may be derived from the derivatives of the potential function, \emph{viz.}
\begin{align}
 \sigma_{i1} &= -\Phi_{i,2}\text{, }\sigma_{i2} = \Phi_{i,1},
 \label{eq:stressfromstressfunction}
\end{align}
and $\sigma_{33}$ from the stress-strain law,
\begin{align}
\sigma_{ij} &= C_{ijkl}u_{k,l},   
\end{align}
for $\epsilon_{33}=0$.

\paragraph{A sharp crack subject to uniform loading}
The general solution for a sharp crack in an anisotropic, homogeneous elastic material loaded by a uniform stress $\sigma^\infty_{ij}$ applied at infinity consists of the uniform stress solution due to $\sigma^\infty_{ij}$ and a disturbed solution due to the presence of the crack~\cite{Ting:1996,Bower:2009,Hwu:2010}. 
We require the singular stress field at the crack tip for the present purpose, i.e., it suffices to focus on the disturbed solution. 

Let the crack of length $2a$ be centrally located at $x_2=0$, $|x_1|<a$. The crack surfaces are assumed traction-free. 
For the disturbed solution, the boundary conditions for the stress function read,
\begin{align}
 {\bm \Phi} &= {\bm 0}\text{ for } |x| = \infty, \nonumber \\
 {\bm \Phi} &= -x_1{\bm t}_\Gamma \text{ for }x_2=\pm 0\text{, }|x_1|<a,\nonumber
\end{align}
that is, the stress vanishes at infinity and a uniform traction ${\bm t}_\Gamma  = [\sigma^\infty_{21}, \sigma^\infty_{22}, \sigma^\infty_{23}]^T$ is applied at the upper crack surface and $-{\bm t}_\Gamma$ is applied at the lower crack surface.
The disturbed solution satisfying these boundary conditions is~\cite{Stroh:1958,Ting:1996}
\begin{subequations}
\begin{align}
 {\bm u }    &= \operatorname{Re}\{ {\bm A} \langle f(z_\alpha) \rangle {\bm B}^{-1} \}{\bm t}_\Gamma,\\ 
 {\bm \Phi } &= \operatorname{Re}\{ {\bm B} \langle f(z_\alpha) \rangle {\bm B}^{-1} \}{\bm t}_\Gamma,
\end{align}
\end{subequations}
where $\langle f(z_\alpha) \rangle$ is the diagonal matrix composed of functions $f(z_\alpha)$,
\begin{subequations}
\begin{align}
 \langle f(z_\alpha) \rangle &= \text{diag}\left[ f(z_1), f(z_2), f(z_3)\right],\\
 f(z_\alpha) &= \sqrt{ z^2_\alpha -a^2 } -z_\alpha.
\end{align}
\end{subequations}
${\bm A}$ and ${\bm B}$ are the normalized Stroh matrices and $p_\alpha$ the Stroh eigenvalues determined from the Stroh formalism, see below. 
$z_\alpha = x_1 + p_\alpha x_2$ is the complex-valued argument  ($\alpha = 1\ldots 3$).
The stresses are given by
\begin{subequations}
\begin{align}
 {\bm t}_1 &\equiv \left[ \sigma_{11}, \sigma_{12}, \sigma_{13}\right]^T = -\operatorname{Re} \{ {\bm B} \langle f_{,2}(z_\alpha) \rangle {\bm B}^{-1} \}{\bm t}_\Gamma,\\
 {\bm t}_2 &\equiv \left[ \sigma_{21}, \sigma_{22}, \sigma_{23}\right]^T = \phantom{-}\operatorname{Re} \{ {\bm B} \langle f_{,1}(z_\alpha) \rangle {\bm B}^{-1} \}{\bm t}_\Gamma,
 \label{eq:general}
\end{align}
\end{subequations}
where the derivatives $f_{,1}$ and $f_{,2}$ read
\begin{subequations}
 \begin{align}
 f_{,1}(z_\alpha) &= \frac{z_\alpha}{\sqrt{z^2_\alpha -a^2}} -1, \\
 f_{,2}(z_\alpha) &= \frac{p_\alpha z_\alpha}{\sqrt{z^2_\alpha -a^2}} -p_\alpha.
\end{align}
\end{subequations}
The stress intensity factors ${\bm K}$ at the crack tips are 
\begin{align}
 {\bm K} &= \sqrt{\pi a}{\bm t}_\Gamma.
 \label{eq:K}
\end{align}

\paragraph{Near crack-tip solution}
Performing a coordinate transformation from ($x_1,x_2,x_3$) to orthonormal cylindrical-polar coordinates ($r$, $\theta$, $x_3$) and shifting the origin to the crack tip at $(a,0,0)$, \emph{viz.}
\begin{align}
z_\alpha= r ( \cos\theta + p_\alpha\sin\theta) + a,
\end{align}
one obtains for the asymptotic behavior of $f_{,1}$ and $f_{,2}$ 
\begin{subequations}
\begin{align}
 f^\text{a}_{,1}(z_\alpha) &= \frac{ \sqrt{a} }{\sqrt{2r(\cos\theta+p_\alpha\sin\theta)}}, \\ 
 f^\text{a}_{,2}(z_\alpha) &= \frac{p_\alpha \sqrt{a} }{\sqrt{2r(\cos\theta+p_\alpha\sin\theta)}}, 
 \label{eq:asymptotic}
\end{align}
\end{subequations}
as $r\to 0$.

\paragraph{Angular dependence of singular stress field}
We seek to determine the $\theta$-dependence of the stress components in cylindrical-polar coordinates $\sigma_{\theta \alpha}$ under pure mode $K_\beta$ loading as $r \to 0$, i.e., $F_{\alpha\beta}(\theta)$-functions in the notation of Rice~\cite{Rice:1992}.
Using the definition of stress intensity factors and Eq.~\eqref{eq:K}, one readily obtains in the asymptotic limit,
\begin{subequations}
\begin{align}
 {\bm t}^\text{a}_1 &= -\operatorname{Re} \{ {\bm B} \langle \frac{p_\alpha}{\sqrt{\cos\theta + p_\alpha \sin\theta}}  \rangle {\bm B}^{-1} \}{\bm K},\\
 {\bm t}^\text{a}_2 &= \phantom{-}\operatorname{Re} \{ {\bm B} \langle \frac{1}{\sqrt{\cos\theta + p_\alpha \sin\theta}} \rangle {\bm B}^{-1} \}{\bm K},
\end{align}
\end{subequations}
from which one may determine the angular dependence of the singular stresses in Cartesian coordinates.
Applying the usual tensor transformation laws, i.e., a rotation through angle $\theta$ around $x_3$, gives the desired singular stresses in cylindrical-polar coordinates.

\paragraph{Stroh's sextic formalism}
We assume that the material is non-degenerate, i.e., possesses six distinct Stroh eigenvalues and a simple or semisimple fundamental elasticity matrix [Eq.~\eqref{eq:fundamentalelasticitymatrix} below].
Let $Q_{ik}$, $R_{ik}$, and $T_{ik}$ be matrices determined from $C_{ijkl}$ for the chosen crack orientation,
\begin{align}
 Q_{ik} &= C_{i1k1}, R_{ik} = C_{i1k2}, T_{ik} = C_{i2k2}, \nonumber
\end{align}
and let further
\begin{align}
{\bm N}_1 &= -{\bm T}^{-1}{\bm R}^{T}, {\bm N}_2 = {\bm T}^{-1}, {\bm N}_3 = {\bm R \bm T}^{-1}{\bm R}^{T}-{\bm Q}.\nonumber
\end{align}
The six right eigenvalues $p$ and corresponding eigenvectors ${\bm \chi}$ of the $6\times 6$, real, and not symmetric fundamental elasticity matrix ${\bm N}$~\cite{Bower:2009},
\begin{subequations}
\begin{align}
 {\bm N} {\bm \chi} &= p{\bm \chi}, \\
 {\bm N} &= \left[ \begin{array}{ll} {\bm N}_1 & {\bm N}_2 \\ {\bm N}_3 & {\bm N}^{T}_1 \\ \end{array} \right], {\bm \chi} = \left[ \begin{array}{l} {\bm a} \\ {\bm b} \\ \end{array} \right],
 \label{eq:fundamentalelasticitymatrix}
\end{align}
\end{subequations}
are the Stroh eigenvalues and Stroh eigenvectors, respectively.
The eigenvalues are complex conjugate pairs $(p, \bar{p})$ with corresponding complex valued eigenvectors $({\bm \chi}, \bar{\bm \chi})$. 
We follow the usual convention that the roots with positive imaginary part are $p_\alpha$, $\alpha = 1\ldots 3$, and the complex conjugates are $p_{\alpha+3} = \bar{p}_\alpha$. 
The traction part ${\bm b}$ of the Stroh vectors are related to the displacement part ${\bm a}$ through
\begin{align}
 {\bm b}_{\alpha} &= \left( {\bm R}^T + p_\alpha {\bm T} \right){\bm a}_{\alpha}, \nonumber \\
 {\bm b}_{\alpha+3} &= \bar{\bm b}_{\alpha}.\nonumber
\end{align}
The ${\bm a }_\alpha$ and ${\bm b}_\alpha$ are collected in the Stroh matrices ${\bm A}$ and ${\bm B}$,
\begin{align}
 {\bm A} &= \left[ {\bm a}_{1}, {\bm a}_{2}, {\bm a}_{3} \right], \quad {\bm B} = \left[ {\bm b}_{1}, {\bm b}_{2}, {\bm b}_{3} \right].\nonumber
\end{align}
Similarly,  $\bar{\bm A}$ and $\bar{\bm B}$ can be defined from $\bar{\bm a}_{\alpha}$ and $\bar{\bm b}_{\alpha}$, respectively.

Let ${\bm \eta}$ be the left eigenvectors of ${\bm N}$, i.e., ${\bm \eta}^T {\bm N} = p {\bm \eta}^T$ and ${\bm \eta}^T = [{\bm b}^T, {\bm a}^T]$. Then, the orthogonality of left and right eigenvectors associated with different eigenvalues suggests the normalization~\cite{Ting:1996,Bower:2009},
\begin{align}
 {\bm \eta}^T_\alpha {\bm \chi}_\beta & = \delta_{\alpha\beta}, \quad \alpha, \beta = 1\ldots 6.\nonumber
\end{align}

The real-valued Barnett-Lothe tensor ${\bm L}$ for non-degenerate materials can be computed from the Stroh matrices~\cite{Bower:2009,Ting:1996}, \emph{viz.}
\begin{align}
 {\bm L} &= -2i{\bm B} {\bm B}^{T} = \operatorname{Re}\{-i{\bm B}{\bm A}^{-1}\} .
\end{align}

\section{\label{sec:appLF}Local-environment dependence of USF energies and surface energies}

The local environment dependence was investigated for the surface energies of the \{100\} and \{110\} terminations and the USF energies of the $\langle 111 \rangle \{110\}$ and $\langle 111 \rangle \{112\}$ slip systems in HfNbTiZr.
All calculations were performed using the EMTO method 
and the super cell approach to represent a random alloy.
Different alloy configurations were simulated by randomly distributing the four chemical species on the lattice (ensuring that the nominal composition is preserved), and 50 configurations were modeled for each surface termination and unstable fault considered.

The surface subsystems in these surface energy calculations were modeled as $4\times 4$ lateral replica of the respective reference cell with one site per layer. The slabs were eight atomic layers thick (generating 128 atomic sites), and the two surfaces in each slab were separated by vacuum corresponding to four layers (simulated by 64 ``empty'' sites), resulting in a total of 192 sites in each simulation cell. Surface relaxation was performed in that all surface sites were relaxed by equal amounts perpendicular to the surface plane.
The total surface areas are $2Na^2$ and $\sqrt{2}Na^2$ for the \{100\} termination and the \{110\} termination, respectively, where $2N=32$ is the number of surface atoms counting both surfaces.
The bulk subsystems were identical to the surface subsystems in their dimensions (without the vacuum layers). Brillouin zone integrations for the surface subsystem and the bulk subsystem were performed on a $2\times 2\times 1$ $k$-point mesh and a $2\times 2\times 2$ $k$-point mesh, respectively.

It should be noted that calculations for the \{111\} surface termination required larger slabs than was computationally feasible and were not carried out.
For instance, assuming a $4\times 4$ lateral replica as for the other two facets, a slab thickness of nine atomic layers  (144 sites) and vacuum corresponding to six layers (96 ``empty'' sites) results in a total of 240 sites in the simulation cell for the surface subsystem. A smaller slab thickness, i.e., six atomic layers and three layers vacuum, did not ensure decoupled surfaces. 

The \{110\} and \{112\} USF energy calculations, performed by the tilted super cell approach,
were carried out for $5\times 4$ lateral replica of the respective reference cell with one site per layer. The thickness of these cells was set to eight atomic layers each, resulting in 160 sites per simulation cell. 
The fault areas are $N/\sqrt{2}a^2$ for the \{110\} slip plane and $\sqrt{3/2}Na^2$ for the \{112\} slip plane. 
In both cases, $N$ equals 20 since there is only one USF per simulation cell. 
The relaxation of the interplanar distances perpendicular to the slip plane was considered in that all sites belonging to the two layers forming the fault were relaxed by equal amounts.
Brillouin zone integrations were performed on $2\times 2\times 2$ $k$-point meshes.

In order to compare the results of these super cell calculations to CPA results on an equal footing, that is, to avoid numerical errors due to different sizes of the simulation cells and $k$-point meshes, 
we recomputed all CPA planar fault energies using the same simulation cells and $k$-point meshes as for the super cell calculations. 
This second set of CPA material parameters for HfNbTiZr, shown in Table~\ref{table:distdata}, differs slightly (-2\,\% for both surface energies, -7\,\% for the \{110\} USF energy, negligible for the \{112\} USF energy) from the values listed in Table~\ref{table:lattice_constants} due to the different computational setups. 
The two setups also result in slightly different $K_{\RN{1}\text{e}}$ and $K_{\RN{1}\text{c}}$ (both scale with $\sqrt{\gamma}$); the largest difference in any of the $K_{\RN{1}\text{e}}/K_{\RN{1}\text{c}}$ ratios amounts to -3.5\,\% and occurred for models 10, 11, and 12 (cf.~values in Tables~\ref{table:deviation} and~\ref{table:distdata}).

\begin{figure}
\begin{tabular}{l@{\qquad}l}
			(a) & (b) \\
            \resizebox{0.4\linewidth}{!}{\includegraphics[]{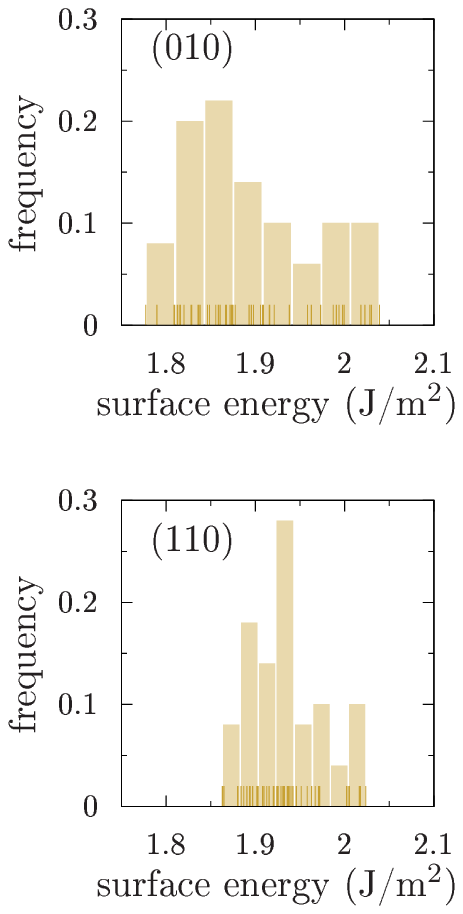}} 	
			&
			\resizebox{0.4\linewidth}{!}{\includegraphics[]{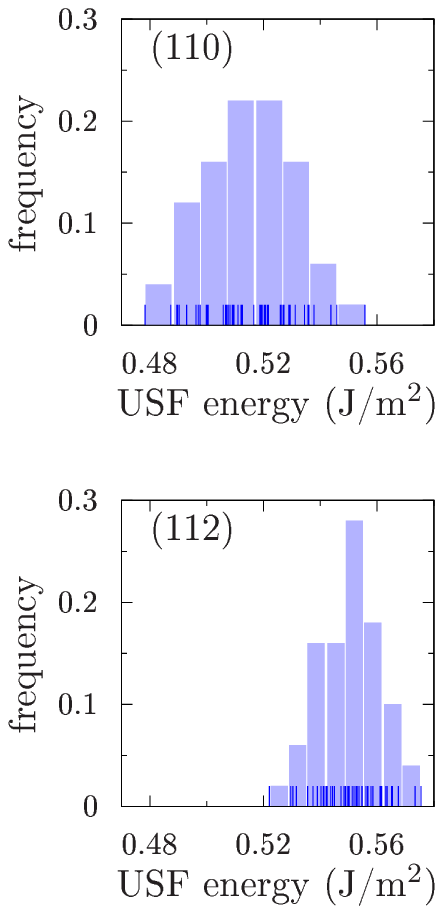}}\\
\end{tabular}
\caption{\label{fig:dist}Distributions (shown as frequency histograms and rug plots) of (a) surface energies and (b) USF energies in HfNbTiZr obtained from super cell calculations over 50 configurations. The distribution averages, standard deviations, minimum and maximum values are specified in Table~\ref{table:distdata}.}
\end{figure}

Figures~\ref{fig:dist}(a) and (b) illustrate narrow distributions of surface energies and USF energies, respectively, over the 50 alloy configurations. 
Table~\ref{table:distdata} lists the distribution averages, standard deviations, minimum and maximum values along with the CPA results and the planar fault areas considered.
Both USF energy distributions are relatively narrower than those of the surface energies as indicated by their standard deviations. Furthermore, 
the ratios of half the difference between the maximum and minimum values to the distribution average amount to 8\,\% and 5\,\% for the $\gamma_{\text{usf},(110)}$  and $\gamma_{\text{usf},(112)}$, respectively. These values also indicate small fluctuations of $\gamma_{\text{usf}}$  around the distribution average.

The surface energy distribution averages are rather similar to their CPA counterparts, i.e., both $\gamma_{\text{s},(010)}$ and $\gamma_{\text{s},(110)}$ differ by $\approx 1$\,\%. 
This justifies the use of CPA \{111\} surface energies in estimating the effect of fluctuations in local alloy composition on the $K_{\RN{1}\text{e}}/ K_{\RN{1}\text{c}}$ ratio in place of \{111\} distribution averages (Sec.~\ref{sec:lee}).
The CPA results for the USF energies are slightly larger than the super cell averages, i.e.,  $\gamma_{\text{usf},(110)}$ by 4\,\% and $\gamma_{\text{usf},(112)}$ by 8\,\%.
Calculations for a larger number of super cell configurations or larger fault area could clarify if this difference is related to restricted ensemble averaging in the super cell approach or the mean-field nature of the CPA.

\begin{table}
\caption{\label{table:distdata}Average, standard deviation (SD), minimum (MIN) and maximum (MAX) values of the surface energy and USF energy distributions for HfNbTiZr shown in Fig.~\ref{fig:dist}, as well as CPA results. The planar fault areas used in these CPA and super cell calculations are specified.}
 \begin{tabular}{l*{6}{c}}
 \toprule
 & & \multicolumn{5}{c}{Planar fault energy (J/m$^2$)}\\
 \cmidrule(lr){3-7}
  Material & Area &  & \multicolumn{4}{c}{Super cell}\\
  \cmidrule(lr){4-7}
  parameter & (\AA$^2$) & CPA & Average & SD & MIN & MAX \\
  \midrule
  $\gamma_{\text{s},(010)}$ & 395 & 1.922 & 1.897 & 0.072 & 1.777 & 2.039 \\
  $\gamma_{\text{s},(110)}$ & 279 & 1.919 & 1.935 & 0.041 & 1.863 & 2.024 \\
    \midrule
  $\gamma_{\text{usf},(110)}$ & 174 & 0.536 & 0.515 & 0.017 & 0.478 & 0.556 \\
  $\gamma_{\text{usf},(112)}$ & 302 & 0.592 & 0.550 & 0.011 & 0.522 & 0.575 \\
  \bottomrule
 \end{tabular}
\end{table}

\section{\label{sec:app2}Deviation from pure mode \RN{1} loading: numerical values}

Table~\ref{table:deviationdetail} lists $K_{\RN{1}\text{e}}$ and $ K_{\RN{1}\text{c}}$ values determined from Eq.~\eqref{eq:KIemixed} and Eq.~\eqref{eq:KIcmixed}, respectively, for mixed-mode loadings. Values for not normalized ${\bm k}$ vectors and normalized ${\bm k}$ vectors ($|\bm{k}|=1$) are given. 
Numerical values for the ratio $K_{\RN{1}\text{e}}/ K_{\RN{1}\text{c}}$ are listed in Table~\ref{table:deviation}. Note that this ratio does not depend on the norm of ${\bm k}$.

\begin{table*}[thb]
\caption{\label{table:deviationdetail}Numerical values for $K_{\RN{1}\text{e}}$ and $ K_{\RN{1}\text{c}}$, given as $\frac{K_{\RN{1}\text{e}}}{K_{\RN{1}\text{c}}}$, for mixed-mode loadings. Values outside/in parentheses refer to unnormalized/normalized ${\bm k}$ vectors (${\bm K} = K_\RN{1} {\bm k}$).}
 \begin{tabular}{ccccc}
 \toprule
           & \multicolumn{2}{c}{HfNbTiZr} & \multicolumn{2}{c}{MoNbTaVW} \\
           \cmidrule(lr){2-3}\cmidrule(lr){4-5}
 ${\bm k}^T$ & Model 5 & Model 11 & Model 5 & Model 11\\
 \midrule
 $[0.00,1,0]$ & $\frac{0.680}{0.658}$ & $\frac{0.796}{0.689}$ & $\frac{1.454}{1.091}$ & $\frac{1.489}{1.177}$\\
 $[0.05,1,0]$ & $\frac{0.659}{0.658}$ $\left(\frac{0.658}{0.658}\right)$ & $\frac{0.841}{0.689}$ $\left(\frac{0.840}{0.689}\right)$ & $\frac{1.414}{1.091}$ $\left(\frac{1.412}{1.090}\right)$ & $\frac{1.568}{1.178}$ $\left(\frac{1.567}{1.176}\right)$ \Tstrut\\
 $[0.10,1,0]$ & $\frac{0.640}{0.658}$ $\left(\frac{0.637}{0.655}\right)$ & $\frac{0.893}{0.689}$ $\left(\frac{0.888}{0.686}\right)$ & $\frac{1.379}{1.091}$ $\left(\frac{1.373}{1.264}\right)$ & $\frac{1.661}{1.177}$ $\left(\frac{1.653}{1.171}\right)$ \Tstrut\\
 \bottomrule
 \end{tabular}
\end{table*}

\end{document}